\documentclass{vgtc}                          

\ifpdf
  \pdfoutput=1\relax                   
  \usepackage{graphicx}                
  \DeclareGraphicsExtensions{.pdf,.png,.jpg,.jpeg} 
\else
  \usepackage{graphicx}                
  \DeclareGraphicsExtensions{.eps}     
\fi%

\graphicspath{{figures/}{pictures/}{images/}{./}} 

\usepackage{microtype}                 
\PassOptionsToPackage{warn}{textcomp}  
\usepackage{textcomp}                  
\usepackage{mathptmx}                  
\usepackage{times}                     
\usepackage{cite}                      
\usepackage{tabu}                      
\usepackage{booktabs}                  
\usepackage{adjustbox}
\usepackage{tabularx} 
\usepackage{array}
\usepackage{caption}
\usepackage{subcaption}
\usepackage{gensymb}
\usepackage{amsmath}
\usepackage{xcolor}
\usepackage{makecell}
\usepackage{algorithm}
\usepackage{algpseudocode}
\usepackage{multirow}
\usepackage{enumitem}
\usepackage{float}
\usepackage{hyperref}
\usepackage{tabularray}
\usepackage{booktabs}  
\newcolumntype{P}[1]{>{\centering\arraybackslash}p{#1}}
\newcolumntype{M}[1]{>{\centering\arraybackslash}m{#1}}
\newcolumntype{C}{>{\centering\arraybackslash}X}

\usepackage{caption}
\usepackage{subcaption}
\usepackage{tabu}
\usepackage{booktabs}
\usepackage{array}
\usepackage{adjustbox}
\usepackage{tabularx}
\usepackage{textcomp}
\usepackage{xcolor}
\usepackage{interval}
\usepackage{enumitem}
\usepackage{algorithm}
\usepackage{algpseudocode}

\onlineid{0}

\vgtccategory{Research}

\vgtcinsertpkg

\title{Adaptive Multi-Resolution Encoding for Interactive Large-Scale Volume Visualization through Functional Approximation}

\author{Jianxin Sun\thanks{e-mail: jianxin.sun@huskers.unl.edu}\\ %
        \scriptsize University of Nebraska-Lincoln %
\and David Lenz\thanks{e-mail: dlenz@anl.gov}\\ %
     \scriptsize Argonne National Laboratory %
\and Hongfeng Yu\thanks{e-mail: yu@cse.unl.edu}\\ %
     \scriptsize University of Nebraska-Lincoln
\and Tom Peterka\thanks{e-mail: tpeterka@mcs.anl.gov}\\ %
     \scriptsize Argonne National Laboratory     
     }

\teaser{
  \centering
  \includegraphics[width=0.8\linewidth]{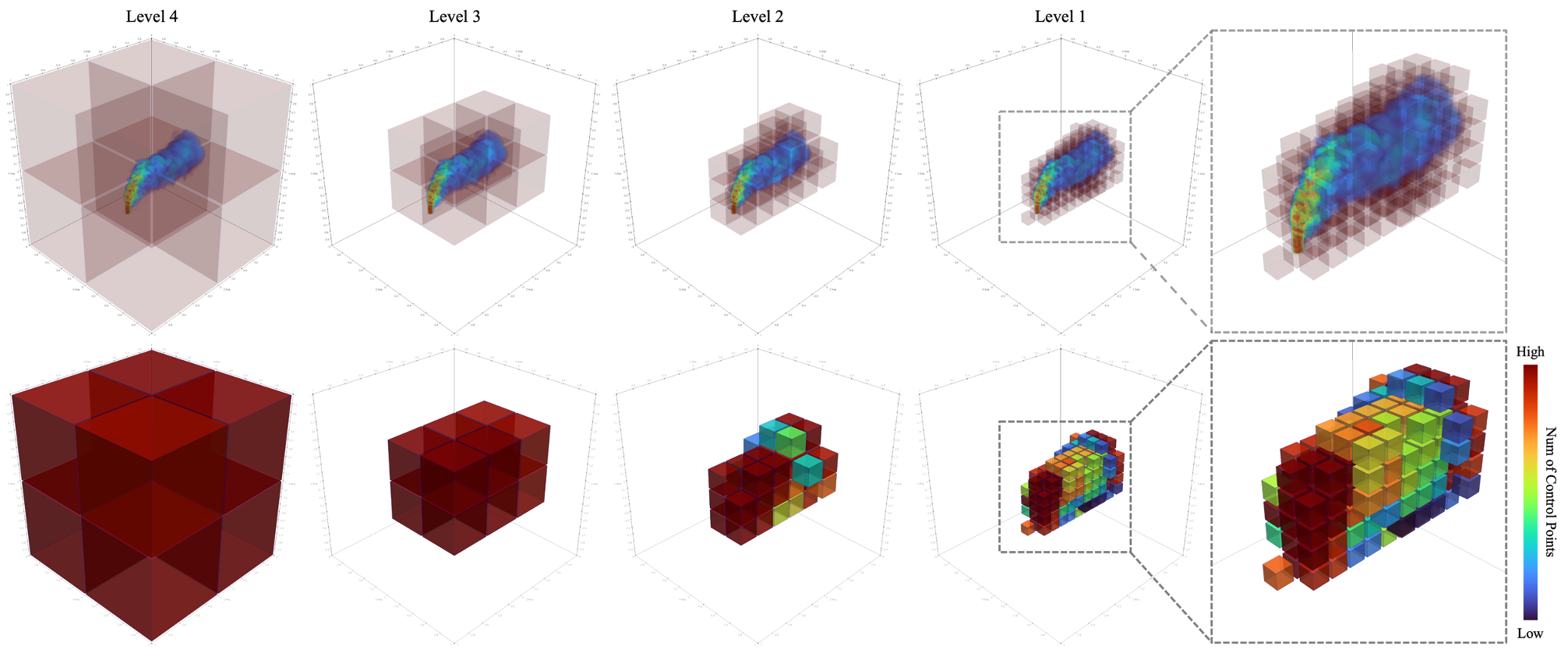}
  \caption{Complex micro-blocks with the dataset. The first row shows the distribution of complex micro-blocks. The second row shows the mapping between the suitable number of control points ($NCP^*$) for each complex micro-block and color.}
  \label{fig:simple_complex_blocks}
}

\abstract{Functional approximation as a high-order continuous representation provides a more accurate value and gradient query compared to the traditional discrete volume representation. Volume visualization directly rendered from functional approximation generates high-quality rendering results without high-order artifacts caused by trilinear interpolations. However, querying an encoded functional approximation is computationally expensive, especially when the input dataset is large, making functional approximation impractical for interactive visualization. In this paper, we proposed a novel functional approximation multi-resolution representation, Adaptive-FAM, which is lightweight and fast to query. We also design a GPU-accelerated out-of-core multi-resolution volume visualization framework that directly utilizes the Adaptive-FAM representation to generate high-quality rendering with interactive responsiveness. Our method can not only dramatically decrease the caching time, one of the main contributors to input latency, but also effectively improve the cache hit rate through prefetching. Our approach significantly outperforms the traditional function approximation method in terms of input latency while maintaining comparable rendering quality.%
} 

\CCScatlist{
  \CCScatTwelve{Functional approximation}{Large-scale data}{Multi-resolution}{Volume visualization}
}

\begin{document}

\firstsection{Introduction}
\maketitle
Interactive volume visualization techniques play important roles in helping researchers from various domains to efficiently discover insightful intrinsic patterns from scientific datasets collected from diverse fields, such as medical imaging, meteorology, materials science, and physical simulations. A visualization system that responds promptly to user interactions, adjusting according to the data or viewing operations, can significantly improve the efficiency and effectiveness of exploring complex scientific datasets. However, the rapid expansion and exponential increase in the size of these datasets pose challenges for interactive systems, particularly on hardware platforms with constrained memory resources and limited I/O bandwidth compared to the size of the data being visualized, resulting in delays in rendering results. 

Researchers have devised a variety of strategies to address this performance challenge. Out-of-core techniques~\cite{8695851} partition the dataset into manageable segments, referred to as \emph{micro-blocks} in this paper, which can be dynamically loaded according to users' exploration, thereby avoiding the necessity of loading the entire dataset. Data streaming strategies~\cite{khan2023web} enable an incremental rendering of a dataset as it becomes available, utilizing both push and pull models. Multi-resolution methods~\cite{6876002}, through a hierarchical data size reduction with multiple levels of detail (LOD)~\cite{Rossignac1993Multiresolution3A, GUTHE200451}, further decrease the size of data sent to the rendering pipeline. Data compression techniques~\cite{son2014data} can be leveraged to minimize the size of datasets produced in scientific research while maintaining crucial information for processing and visualization.  
However, these approaches often involve a trade-off between rendering quality (via methods like multi-resolution techniques and data compression) and input responsiveness (such as out-of-core and streaming methods). It remains challenging to develop techniques capable of achieving a more comprehensive balance across these critical aspects.

Recent works in multivariate functional approximation-based volume rendering~\cite{peterka_ldav18, sun2023mfa, 10386434} provide high-quality results from a compact continuous model, which is suitable for visualizing very large-scale datasets with high quality. Functional data analysis~\cite{ramsay2002applied, ferraty2006nonparametric} employs functional approximation to represent the original discrete dataset with a continuous model, allowing for querying of off-grid values and gradients with increased accuracy via the computation of geometric bases. Common choices of basis functions used for functional analysis are Fourier~\cite{boashash2015time}, wavelet~\cite{jansen2005second}, and geometric splines~\cite{de1978practical} bases.
However, the challenge of employing functional approximation lies in two computational aspects. First, the data needs to be encoded or prefiltered~\cite{8130306, 875199} into continuous models before conducting high-order queries, and this process is extremely computationally expensive when encoding large-scale datasets. Second, once the data is encoded, the model query latency is relatively higher compared to simple filters, especially when the encoded model is large, posing a challenge for real-time applications like interactive visualization systems.

To address these challenges, we proposed a novel functional approximation multi-resolution representation, Adaptive-FAM, through efficient modeling of a large number of data segments with various levels of detail. Adptive-FAM is a compact data representation supporting random access with multiple levels of resolution, addressing the difficulties of large-scale volume rendering. We also implement an out-of-core multi-resolution volume visualization pipeline that parallelizes the direct decoding from Adaptive-FAM through CUDA on GPU for a faster rendering time. We further improve the performance of the proposed pipeline by leveraging prefetching, which decreases the cache miss rate, thereby saving caching time, one of the main contributors to total input latency. Our solution decreases the overall input latency of visualizing a large-scale dataset using functional approximation. The main contributions of this work include:
\setlist{nolistsep}
\begin{itemize}[leftmargin=*]
  \item A novel functional approximation multi-resolution representation, Adaptive-FAM, efficiently encoded from a large number of data segments with various levels of detail.
  \item An out-of-core GPU-accelerated multi-resolution rendering framework that generates high-quality visualization results through parallel decoding of Adaptive-FAM.
  \item A comprehensive investigation of the quality and performance of the proposed method compared with the traditional functional approximation encoder and the state-of-the-art multi-resolution-based volume visualization framework.
\end{itemize}
\section{Background and related works}
\subsection{Out-of-core Methods}
Out-of-core algorithms, also known as external memory algorithms, play an important role in large-scale scientific data visualization including tasks like I/O-efficient volume rendering, isosurface computation, and streamline computation~\cite{https://doi.org/10.1111/cgf.12605, https://doi.org/10.1111/cgf.13671}. Multi-resolution \cite{6876002, https://doi.org/10.1111/cgf.12102} is one of the most popular out-of-core methods for visualizing large-scale volumetric datasets. Multi-resolution techniques take into account the distance between the camera position and each data segment in each view frame and selectively load data segments with varying levels of detail (LOD), resulting in a smaller amount of data being loaded for rendering while still maintaining a similar level of rendering quality. Caching~\cite{qin2020making} is normally integrated with out-of-core visualization pipelines to minimize the times of data movements from slower memory to faster memory in a multi-level memory hierarchical system. Prefetching~\cite{10549835, 7965175} can further improve the Caching performance by preloading the data of interest in advance. Employing an optimized disk data layout~\cite{10.1145/582034.582036} or a pre-calculated lookup table~\cite{10.5555/502125.502134} can enhance the efficiency of accessing raw data in real-time visualization techniques such as progressive slicing and particle tracing. Our rendering framework utilizes similar ideas to dynamically load visible data segments as needed.

\begin{figure}[t]
    \centering 
    \includegraphics[trim=0 25 0 0,clip,width=0.95\columnwidth]{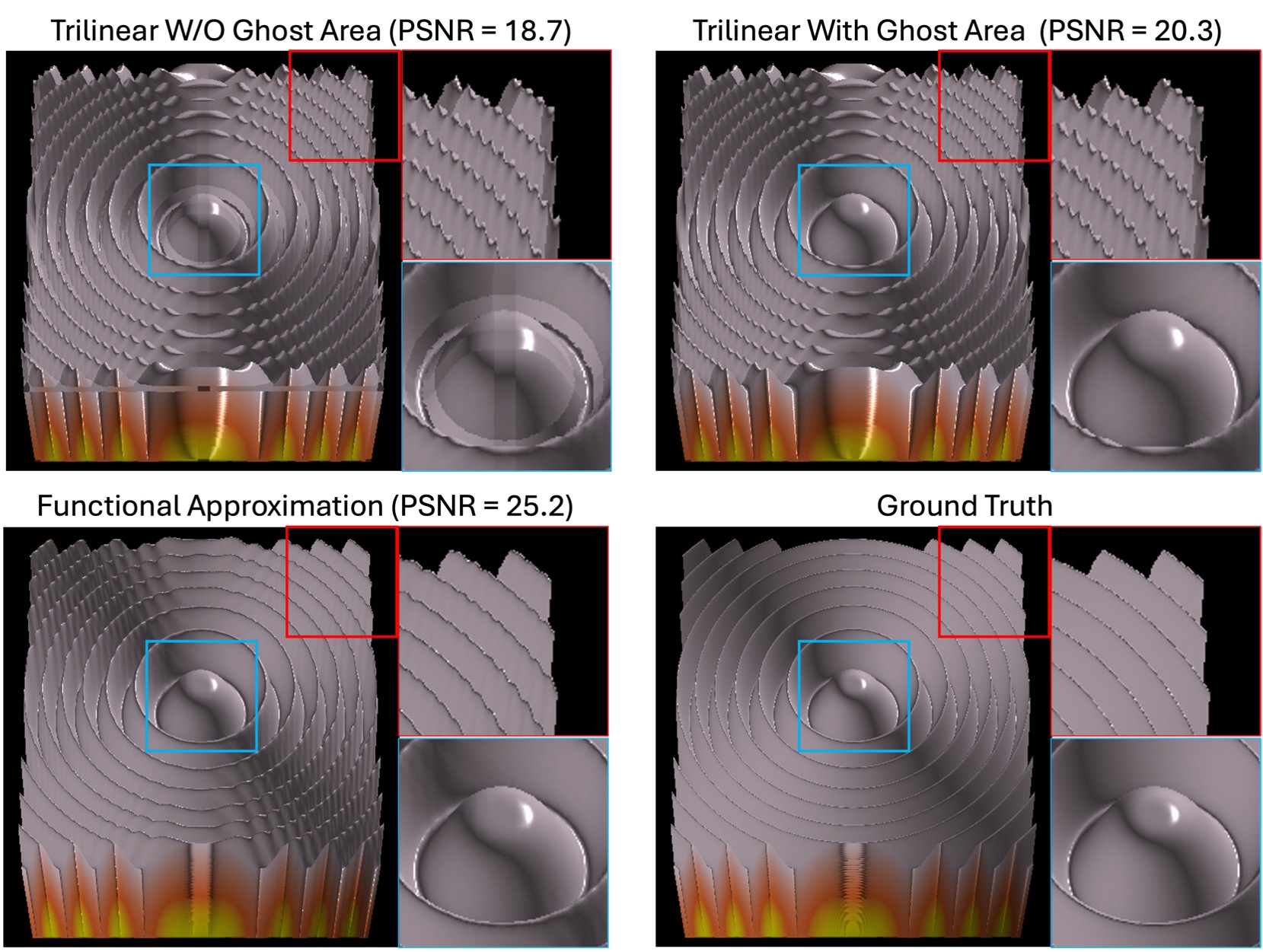}
    \caption{Volume visualization using functional approximation and trilinear interpolation with and without ghost area for parallel rendering in volume space. Input is a Marschner-Lobb dataset with a resolution of $61\times61\times61$. The volume is evenly partitioned into 8 octant blocks.}
    \label{fig:faVstrilinear}
\end{figure}

\subsection{Acceleration Techniques}
Acceleration of visualizing large datasets can be achieved by harnessing environments and platforms such as multi-core CPUs, GPUs, and high-performance computing (HPC) clusters~\cite{zhang2005survey}. Chiang et al.~\cite{964405} create a unified infrastructure to enable parallel out-of-core visualization of unstructured grids. Piringer et al.~\cite{5290719} introduce a versatile multithreaded visualization architecture designed to mitigate challenges associated with multithreading while providing visual feedback. Liu et al.~\cite{Liu2015OnVL} create POIViz, employing a radial representation to parallelize the computation of a 2D layout for multidimensional datasets on both CPU and GPU. Sun et al.~\cite{10386434} proposed a distributed computing method to enhance the interactive performance of visualizing a large-scale model encoded using function approximation with superior quality compared with traditional interpolation methods. However, HPC resources are not easily accessible to non-professional users. Our method tries to offer a solution by enabling high-quality rendering of large-scale data on commercial GPUs at the edge.  

\subsection{Deep Learning-based Volume Visualization}
In recent years, deep learning-based rendering synthesis~\cite{lochmann2016real} has emerged to enhance the performance of rendering volumetric datasets across spatial and temporal domains. Berger et al.~\cite{8316963} use a generative adversarial network (GAN) framework to directly generate the visualization image from view parameters and transfer functions. Weiss et al.~\cite{8918030} and Bauer et al.~\cite{9903564} utilize reconstruction neural networks to speed up the rendering by inferencing the missing detail from sparsely rendered images through inpainting. However, rendering synthesis methods require the creation of comprehensive input and label pairs to train the network, accounting for extensive combinations between view- and data-dependent operations. For large-scale data, preparing these training datasets demands a significant amount of computational resources. Furthermore, the training time is typically lengthy to achieve accurate inference results. Additionally, the memory footprint of rendering synthesis is usually substantial to generate the resulting image in a single shot. Compressing methods leveraging implicit neural representation \cite{lu2021compressive, tang2020deep} are proposed to decrease the network size and optimize I/O intensive operations. Although random access allows querying values or gradients directly from the neural representation without decompression, the training time remains inevitably long even with powerful GPUs when handling complex large-scale scientific datasets~\cite{TANG2024103874, 10175377}. Existing deep learning-based methods prioritize enhancing compression ratios and rendering performance. Sporadic efforts are dedicated to improving rendering quality through non-linear interpolation at off-grid volumetric locations~\cite{zhou2022analysis}. The rendering frameworks we seek to optimize are multi-resolution-based renderers using function approximation, where compression is selectively applied to data partitions without extensive training or modeling of the entire dataset.

\section{Functional Approximation}
In this section, we first explain the advantage of adopting functional approximation to large-scale volume rendering in terms of rendering quality. Secondly, we discuss the main performance issues of rendering a large-scale functional approximation model.

\begin{figure}[t]
    \centering 
    \includegraphics[trim=0 0 0 0,clip,width=0.8\columnwidth]{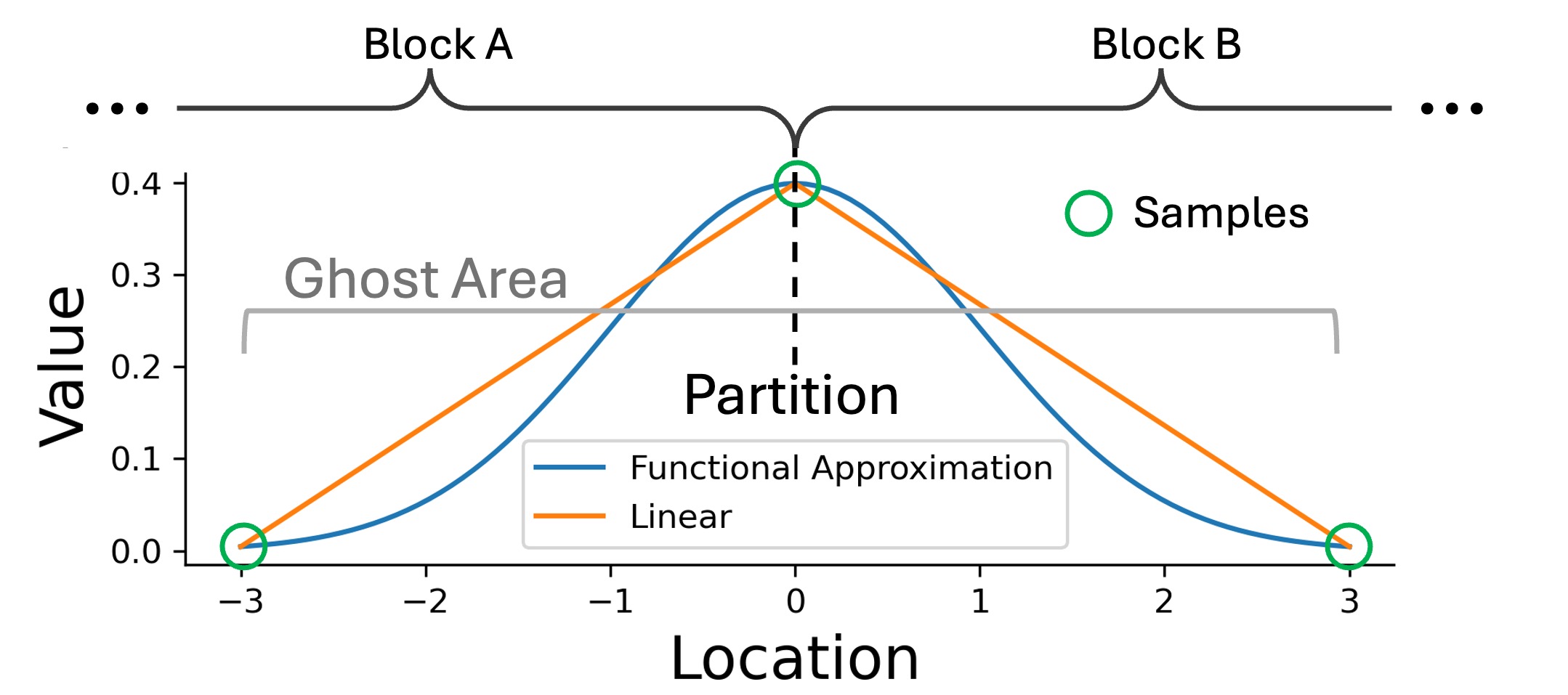}
    \caption{Volume partition with ghost area.}
    \label{fig:between}
\end{figure}

\subsection{Advantages}
\subsubsection{Rendering Quality}
Typical scientific datasets are generally in the form of structured or unstructured grids, and their local domains feature non-linear spatial correlations. The accuracy of interpolation while querying off-grid sample location determines the accuracy of the downstream visualization. Continuous representation of volume data using functional approximation provides a more accurate evaluation of off-grid sample locations than traditional trilinear interpolation when handling scientific datasets. Its control parameters, like the number of control points and polynomial degree, are flexible to adjust for a minimal global error. Detailed rendering quality comparison between functional approximation and other local filters can be found in ~\cite{sun2023mfa}. \autoref{fig:faVstrilinear} shows an example of rendering results using functional approximation and trilinear interpolation. Under the same model size, the functional approximation rendering result is closer to the ground truth rendering of typical scientific datasets.


\subsubsection{Boundary Continuity}
A common practice of speeding up the rendering is to distribute the workload to multiple computing units for parallel processing, which requires the partition of data volume. Multi-resolution methods also need to partition the data volume into data segments, or micro-blocks, and encode each segment with various levels of detail. In order to avoid rendering artifacts in the boundary area, both $C0$ and $C1$ continuities need to be maintained. This is because $C0$ keeps the smoothness across value and $C1$ keeps the smoothness across the shading effect. $C0$ continuity can be easily maintained by partitioning the data segments from shared sample points. In order to be able to calculate the gradient on the boundary, the linear method needs to include a ghost area beyond the boundary for calculating the derivative through central difference as demonstrated in \autoref{fig:between}. Otherwise, lighting artifacts will be observed in the boundary area as shown in \autoref{fig:faVstrilinear}. Ghost areas increase the size of each data segment. For micro-blocks of size $n^3$ with a ghost area of 1 voxel, the ratio of this ghost area to the micro-blocks has a space complexity of $O(1/n)$. Ghost area will take significant data size for micro-block with higher LOD where $n$ is small. If the input volume has a dimension of $n\times n\times n$ and the number of partitions on each edge is $m$, then the total sample needs after the partition will be $k$ times the total samples in the original dataset, where $k$ can be expressed as:
\begin{equation}
k=\frac{(\frac{n}{m} + 1)^3\cdot m^{3}}{n^{3}}=\frac{(n+m)^3}{n^3}, \quad m\subset [1, 2,...n]
\end{equation}
For instance, partitioning 4 segments on each edge of a volume with a resolution of $16\times 16\times 16$ will increase the total micro-block size to about 2 times the input data. Adding a ghost area will greatly increase the total data size to handle for the rendering pipeline, which poses a significant overhead for distributed visualization. In contrast, considering the strong spatial correlation property of general scientific datasets, the functional approximation-encoded mirco-blocks do not require a ghost area to provide a gradient on the boundary close to the C1 continuity because of their higher-order interpolations~\cite{10386434}. This can be observed from the fitted curve in \autoref{fig:between}. The functional approximation is more suitable for modeling data segments considering both data size and modeling accuracy.

\subsection{Opportunities}
Although functional approximation is highly effective at in situ modeling of scientific data with more accurate evaluations than many local filters, its query time becomes the main performance bottleneck for visualizing large-scale datasets\cite{sun2023mfa}. Moreover, the encoding time is also impractical for large-scale multi-resolution modeling. The main goal of this work is to minimize both the encoding time and the overall input latency of rendering large-scale volume datasets modeled by function approximation.
\section{Method}
\subsection{Problem Formulation and Rationale}


\begin{figure}[t]
 \centering 
 \includegraphics[width=\columnwidth]{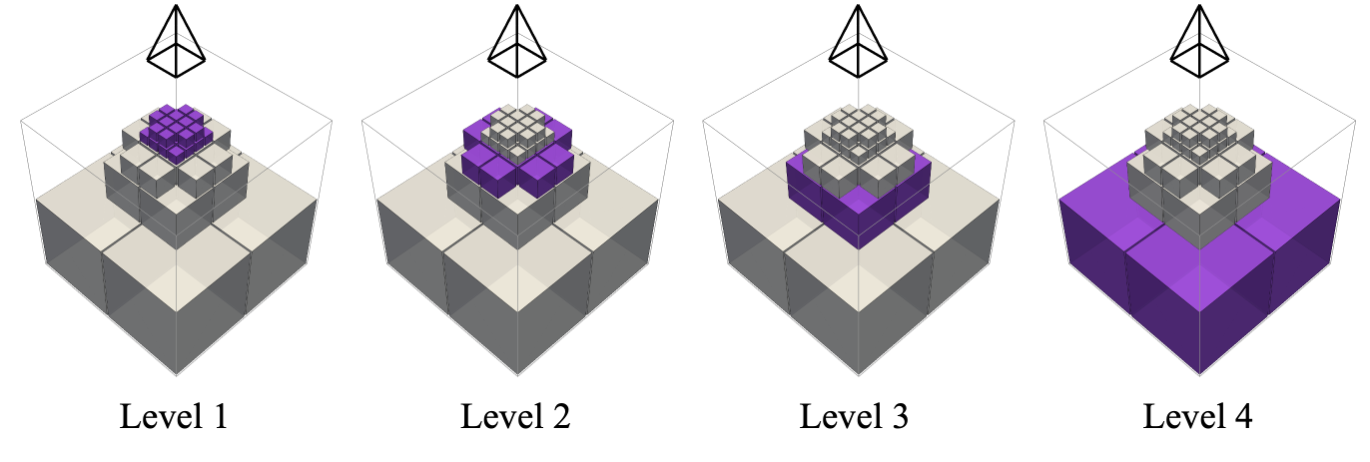}
 \caption{An illustration of micro-blocks at 4 LODs used for rendering from a particular POV. Micro-blocks at Level 1 exhibit the highest LOD, whereas those at Level 4 display the lowest LOD.}
 \label{fig:4_levels}
\end{figure}

\begin{figure}[t]
 \centering 
 \includegraphics[width=\columnwidth]{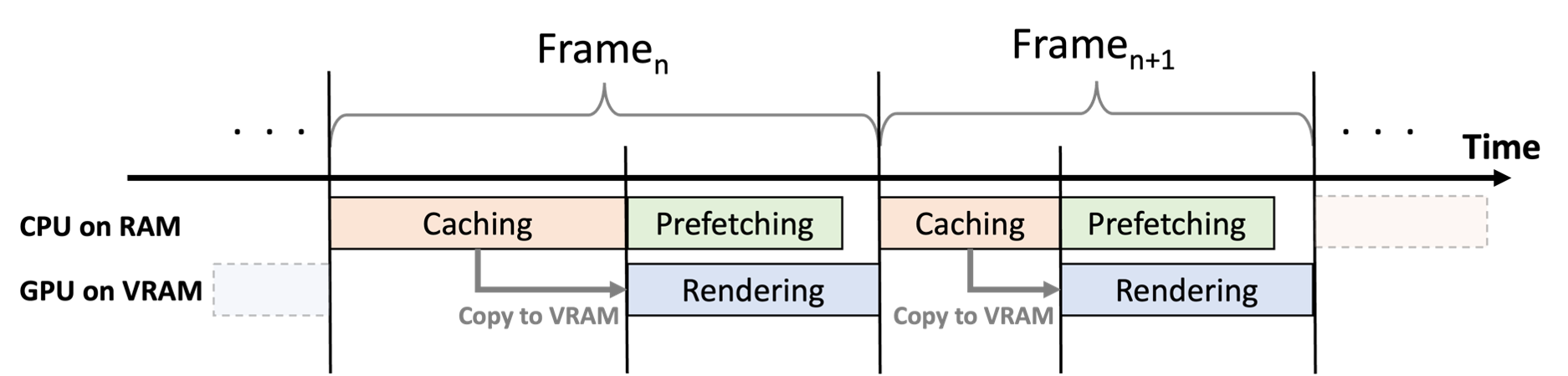}
 \caption{Time breakdown of key processes for generating frames.}
 \label{fig:timing}
\end{figure}

User exploratory behavior toward a dataset can be treated as a sequence of normalized 3D vectors, where each vector's location is the user's point of view (POV) and its direction is the user's direction of view (DOV). For a user exploration with $m$ number of POVs, $\{P_0, P_1, ... P_n, ... P_{m-1}\}$ forms a 3D trajectory of user exploration, where $P_n$ is the $n$th POV. An interactive volume visualization pipeline comprises a sequence of rendering function calls, considering each point of view and its relation to the input dataset. For a specific POV $P_n$, the renderer generates a visualization image frame, $Frame_n$, from the visible region of the dataset, which needs to be loaded to the system memory from storage. Such visible region can be further represented as a group of blocks (micro-blocks) with various levels of detail (LOD). Lower resolution content (i.e., low LOD) is used for regions farther from the user's POV to decrease the overall data size for loading. \autoref{fig:4_levels} highlights micro-blocks for each LOD used to generate a specific frame. \autoref{fig:timing} illustrates the key processes of a typical GPU-accelerated multi-resolution volume rendering pipeline with prefetching. There are three key operations: 
\begin{itemize}[leftmargin=*]
  \item The caching operation involves loading all visible blocks of the current POV from the storage to the system cache based on a certain replacement policy. Subsequently, the cached data is copied to the VRAM on the GPU for rendering.
  \item The rendering operation produces the visualization image based on the current settings derived from the user's data operations and viewing preferences.
  \item The prefetching operation predicts the visible micro-blocks for the next POV using prefetching algorithms and concurrently updates the cache while rendering. To reduce the overall input latency, prefetching needs to cease once rendering is completed.
\end{itemize}

Both caching and prefetching operations are executed on the CPU, while the rendering is accelerated by running on the GPU. The input latency for a specific POV is determined by the combined duration of caching and rendering. The caching time is influenced by factors such as the size of micro-blocks and the caching miss rate. The rendering time depends on the complexity of the rendering algorithm and the capability of rendering hardware. Given a user's exploratory trajectory, the performance objective of our method is to minimize both the caching time and rendering time to achieve a responsive input latency.

The rationale behind our proposed idea stems from the following observations.
First, scientific data typically exhibit strong spatial correlation, often maintaining at least C1 continuity within the 3D domain. By utilizing functional approximation to model two contiguous micro-blocks of the same LOD, the decoded values from the models around their boundary not only demonstrate C0 continuity but also approximate C1 continuity. This approach contributes to saving model size by encoding micro-blocks without the need for a ghost area to support gradient queries. We refer to the functional approximation model from a micro-block as a \emph{micro-model}. Second, standard scientific datasets often encompass large backgrounds or surrounding regions where values remain generally constant, while the central areas represent the research focus with more dynamic values. These datasets mainly originate from domains such as medicine, biology, scientific simulations, and 3D object scans. In this work, we name regions with nearly constant values as \emph{simple regions} and those with more dynamic values as \emph{complex regions}. Therefore, employing lower resolutions to model simple regions generally would not compromise accuracy, allowing for significantly smaller model sizes. This strategy substantially reduces caching time when loading micro-models of such regions. This is our initial effort at minimizing input latency by reducing caching time through minimizing the size of micro-models to be loaded. Consequently, the time saved during caching can offset a portion of the time used during rendering. Additionally, we design a parallel functional approximation-based multi-resolution volume rendering framework with an efficient decoding tailored for large-scale datasets.

\subsection{Adaptive Encoding Algorithms}
\begin{figure}[t]
    \centering 
    \includegraphics[trim=0 0 0 0,clip,width=\columnwidth]{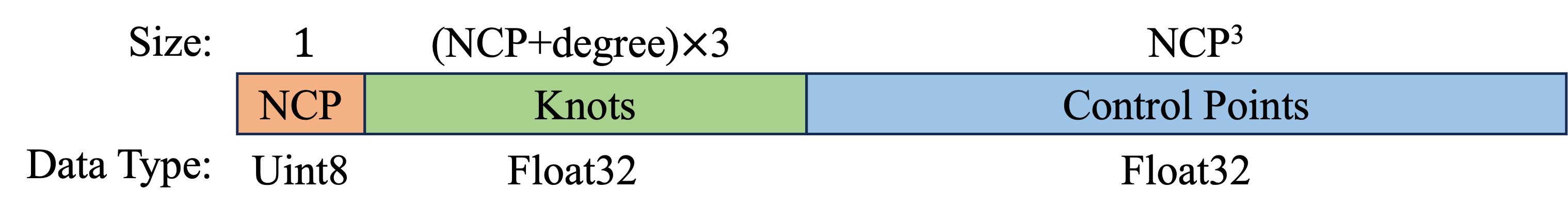}
    \caption{Micro-model storage layout. Knot and Control Point determine the shape and properties of the spline.}
    \label{fig:micro-model}
\end{figure}

\begin{figure*}[t]
 \centering 
 \includegraphics[trim=0 0 0 0,clip,width=0.9\linewidth]{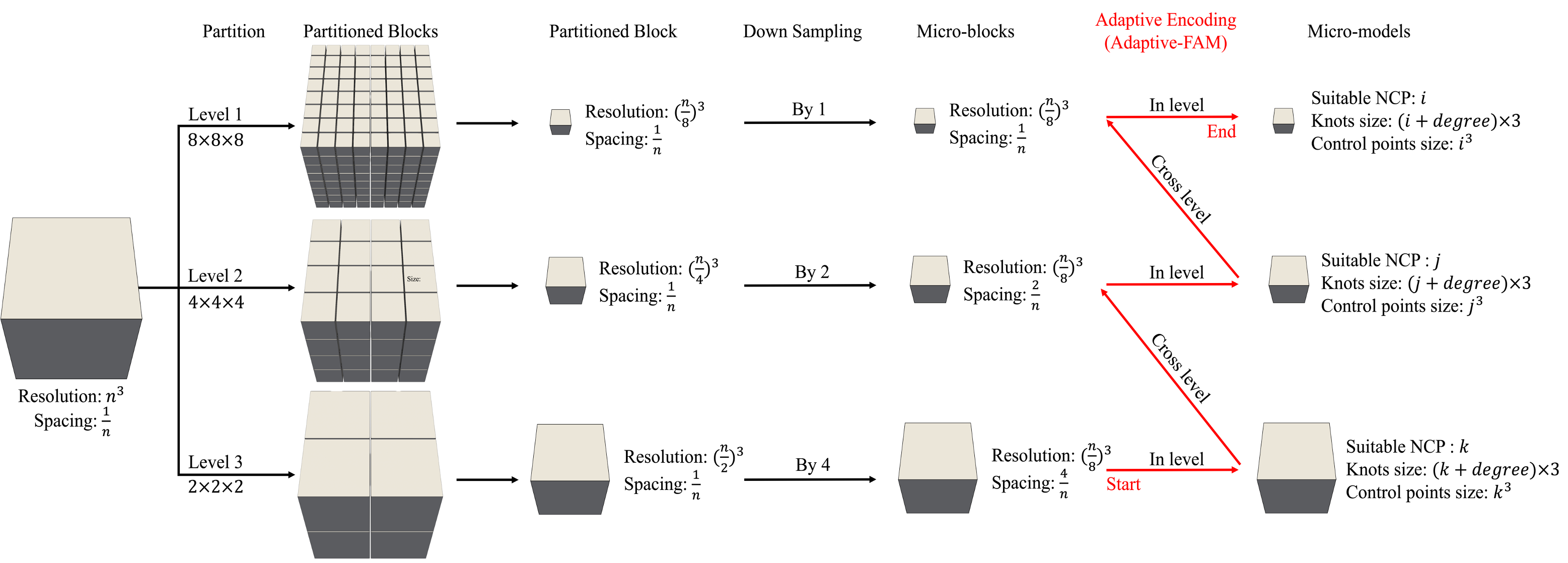}
 \caption{Encoding pipeline for generating micro-models of 3 LODs from input data using proposed adaptive encoding.}
 \label{fig:encoding}
\end{figure*}

For the functional approximation encoder, we have opted for multivariate functional approximation (MFA)~\cite{peterka_ldav18}, which provides the ability to random access for both value and derivative anywhere in the spatial domain with high-order accuracy. 
The encoded model's size can be compressed by adjusting one of its control parameters, the number of control points (NCP), which serves as the primary determinant of the size of encoded micro-models. The storage layout of our encoded micro-model is shown in \autoref{fig:micro-model} where the file size (in bytes) of the micro-model can be calculated as:
\begin{equation}
Size(Micro\textendash model)=1 + ((NCP + degree) \times 3 + NCP^3) \times 4
\end{equation}
Where the degree is another encoding control parameter determining the polynomial order of the fitted B-spline. Except for model size, NCP also determines the fitting accuracy of the functional approximation. In general, a larger NCP gives higher fitting accuracy, but the optimal NCP for a given error bound is data-dependent. Finding a minimal NCP under a specific error involves an exhaustive search. For a 3D dataset with a resolution of $n^3$, MFA encoding for all possible NCP values has a time complexity of $O(n^4)$~\cite{peterka_ldav18} and the total time complexity of encoding all partitioned micro-blocks for a multi-resolution framework with $m$ LODs is:
\begin{equation}
\sum_{k=1}^{m} (2^k)^3 \cdot O((\frac{n}{k})^4) = O(8^m \cdot n^4),
\end{equation}
where the number of micro-blocks increases twofold from the $i$th LOD to the $i+1$th LOD along each $x$, $y$, and $z$ direction. When dealing with a large-scale dataset (i.e., a large $n$) with multiple LODs (i.e., a large $m$), the encoding time can become impractical. To address this challenge, we propose the adaptive functional approximation multi-resolution encoding method designed to efficiently encode micro-blocks into micro-models. The Adaptive-FAM data representation is composed of a collection of micro-models encoded with optimized NCP for a given error bound. \autoref{fig:encoding} demonstrates an example of the encoding pipeline for generating micro-models from input data for three LODs (LOD goes up from level 3 to 1) using our proposed adaptive encoding. For the input data with a resolution of $n^3$, it is first partitioned into blocks for each LOD. In order to secure the C0 continuity, the neighboring blocks partitioned share one sample on the boundary in each $x$, $y$, and $z$ direction. Then, each portioned block is downsampled into micro-blocks according to the resolution requirement of each LOD. Finally, each micro-block is efficiently encoded into a micro-model using our adaptive encoding that incorporates two levels of encoding, in-level and cross-level.

\subsubsection{In-level Adaptive Encoding}
\begin{figure}[t]
    \centering 
    \includegraphics[trim=0 0 0 0,clip,width=0.7\columnwidth]{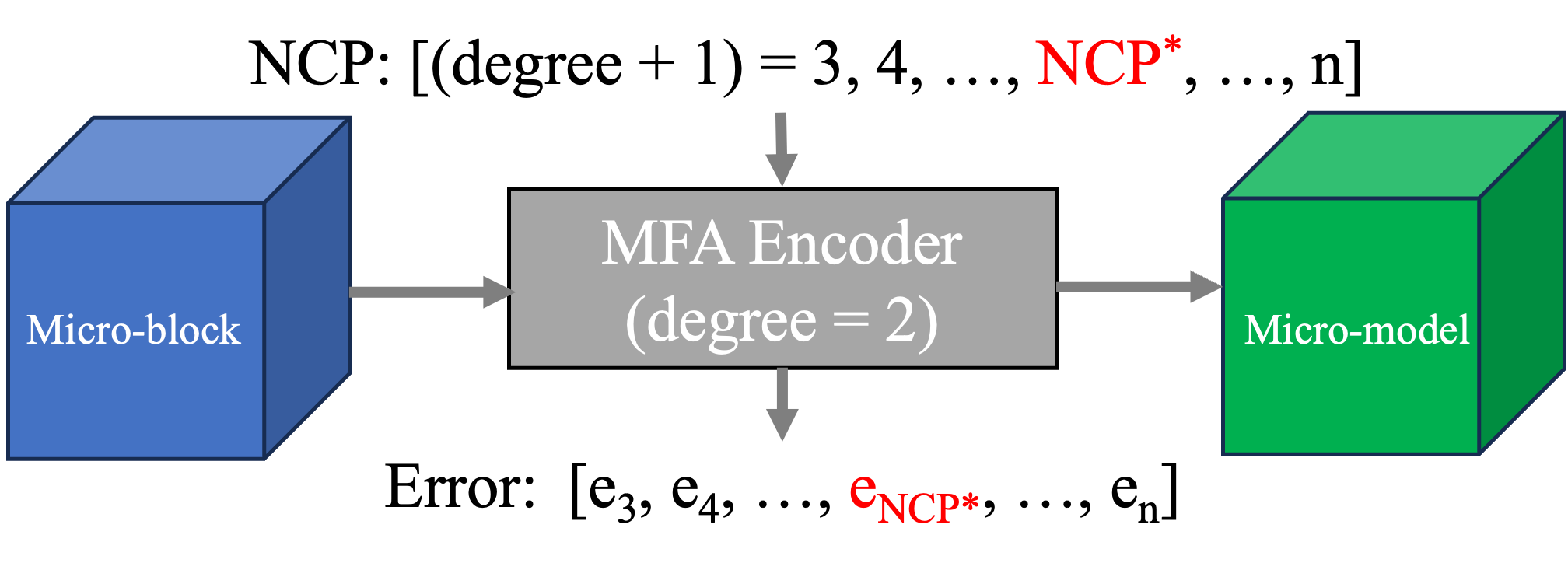}
    \caption{In-level adaptive encoding using MFA encoder.}
    \label{fig:exhaust}
\end{figure}

In-level adaptive encoding conducts an exhaustive search to find the suitable NCP for a required error bound. For a micro-block with a resolution of $n^3$, the encoder uses each valid NCP ($NCP \in [degree + 1, n]$) and records corresponding error metrics to construct a map $Error(NCP)$, as shown in \autoref{fig:exhaust}. For a required error, $e$, the suitable $NCP^*$ for a specific micro-block is determined by:

\begin{equation}
NCP^*=\arg \min_{NCP} (Error(NCP) < e)
\end{equation}
A larger NCP will normally encode a model with lower error, so the checking can start from the highest possible NCP which is $n$.

\subsubsection{Cross-level Adaptive Encoding}
The goal of cross-level adaptive encoding is to avoid the exhaustive search of the in-level adaptive encoding for finding the suitable NCP. Since the level with a lower LOD has fewer micro-blocks to encode from, the idea is to start the in-level adaptive encoding from the level with the lowest LOD. Then we can reuse the result of suitable NCP from the current level for encoding the levels with higher LOD. After in-level adaptive encoding of a LOD, we define the micro-blocks with a suitable NCP equaling to the lowest possible NCP value, $degree + 2$, as simple micro-blocks, while the rest of micro-blocks as complex micro-blocks. Simple micro-blocks are from regions with constant values or values with very small deviations. Complex micro-blocks are from regions with more dynamic values. The first row of \autoref{fig:simple_complex_blocks} shows only the complex micro-blocks overlapped with the dataset for each LOD. We can see that the group of complex micro-blocks from the lower LOD will always cover the group of complex micro-blocks from the higher LOD. From lowest LOD to highest LOD, the process of cross-level adaptive encoding is to only execute the expensive in-level adaptive encoding to the micro-blocks covered by the complex micro-blocks of the previous LOD. This encoding path is highlighted as a red path in \autoref{fig:encoding}. The second row of \autoref{fig:simple_complex_blocks} shows the in-level adaptive encoding result of complex micro-blocks of each LOD with a color map on their suitable NCP values. We can observe that the complex micro-blocks are not only located at the region of the dataset but also reflect the complexity of the underlying data through their suitable NCP results. In this way, we minimize the micro-model sizes encoded from simple micro-blocks while still capturing the feature-riched complex micro-blocks with micro-models with adequate NCPs. In the meantime, we avoid unnecessary encoding for all the simple micro-blocks across all levels of detail. Algorithm \ref{alg:adaptive-FAM} lists the detailed steps of performing our adaptive encoding.

\begin{algorithm}[h]
\caption{Adaptive Encoding}\label{alg:adaptive-FAM}
\textbf{Input:} micro-block ($n^3$) of all LODs, Error Bound\\
\textbf{Output:} micro-models of all LODs.
\begin{algorithmic}[1]
\State // Do in-level adaptive encoding on the lowest LOD
\For{ each micro-block in $LOD_{low}$ }
    \State errs = []
    \For {each NCP in $[n, degree + 2]$}
        \State err = model(NCP, degree)
        \State push err to errs
        \If{err $>$ Error Bound}
            \State push micro-block to $ComplexSet_{LOD_{low}}$
        \EndIf
    \EndFor
    \State Find $NCP^*$ from errs to encode the micro-block
\EndFor
\State // Do cross-level adaptive encoding
\For {each LOD in $[LOD_{low} + 1, LOD_{high}]$}
    \For{ each micro-block in LOD}
        \If{ micro-block is covered $ComplexSet_{LOD - 1}$}
            \State errs = []
            \For {each $NCP$ in $[n, degree + 2]$}
                \State err = model(NCP, degree)
                \State push err to errs
                \If{err $>$ Error Bound}
                    \State push micro-block to $ComplexSet_{LOD}$
                \EndIf
            \EndFor
            \State Find $NCP^*$ from errs to encode the micro-block
        \Else
            \State Use $NCP*$ = degree + 1 to encode the micro-block
        \EndIf
    \EndFor
\EndFor
\end{algorithmic}
\end{algorithm}


\subsection{Multi-Resolution Framework}
\begin{figure}[t]
    \centering 
    \includegraphics[width=0.95\columnwidth]{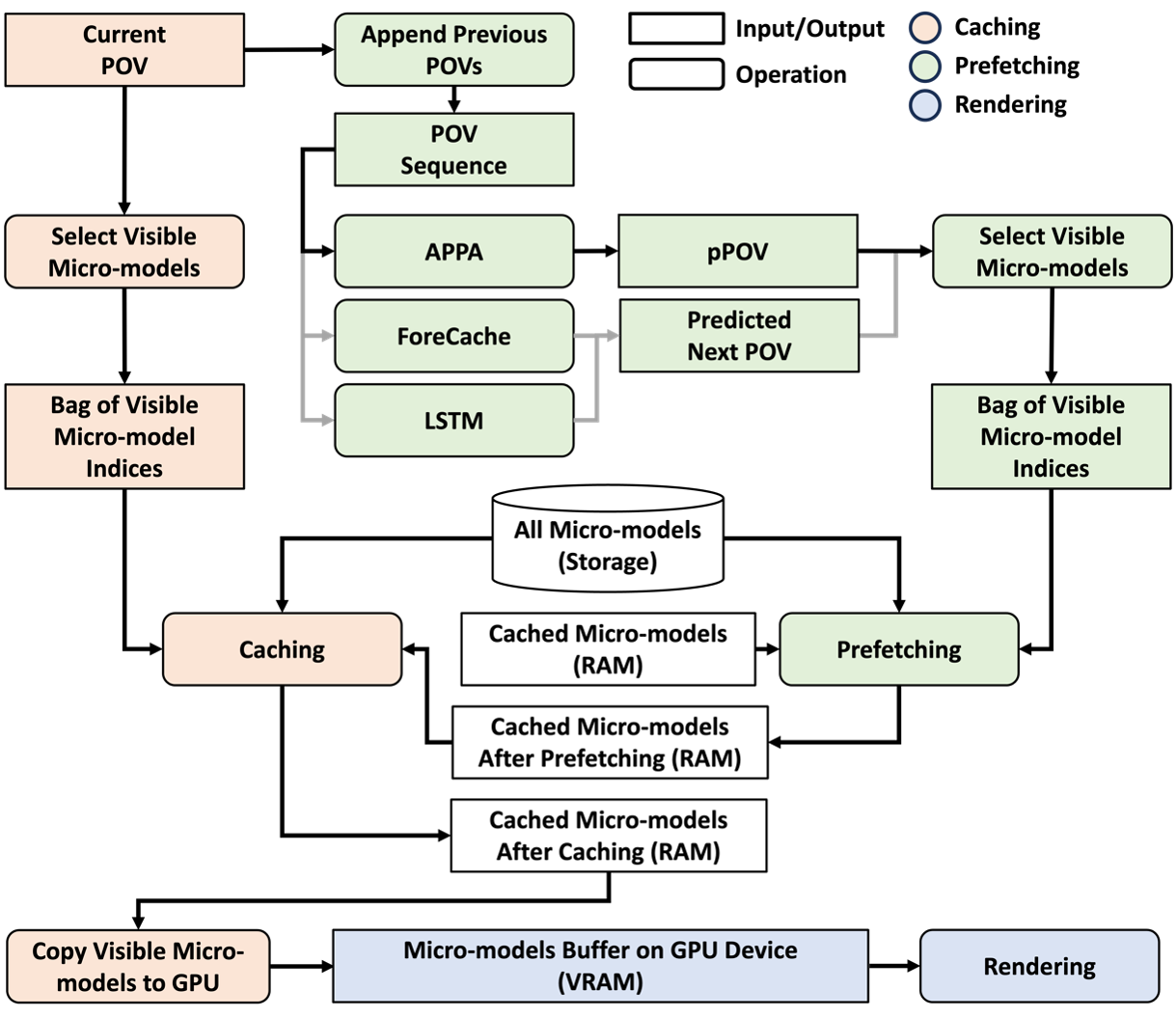}
    \caption{Proposed out-of-core GPU-accelerated multi-resolution rendering framework with prefetching (APPA is selected).}
    \label{fig:pipeline}
\end{figure}

In this section, we present our out-of-core GPU-accelerated multi-resolution rendering framework that generates high-quality visualization results through directly decoding Adaptive-FAM micro-models. \autoref{fig:pipeline} shows the detailed data flow and operations of the proposed framework. There are three main procedures which are caching, prefetching, and rendering.

\subsubsection{Caching}
Caching utilizes spatial locality to decrease micro-model movement frequency across the memory hierarchy. Caching involves three sub-steps as shown in Algorithm \ref{alg:caching}: 1) Find all visible micro-models for the current user POV and collect the indices of micro-models with various LOD. 2) Update the cache in system RAM by loading the missing micro-models from storage according to the LRU replacement policy. It is worth noticing that the cache updated by the caching procedure is the cache after prefetching which happens during the rendering of the previous POV. 3) Copy the data of visible micro-models from system memory to GPU for rendering. The caching time considered in the experiment section is the sum of those 3 sub-steps. In our framework, all the caching-related operations are handled by a CPU thread. 

\begin{algorithm}[t]
\caption{Caching}\label{alg:caching}
\textbf{Input:} $POV_n$.\\
\textbf{Output:} Cached micro-models for $POV_n$.
\begin{algorithmic}[1]
\State $rendering\_done = false$
\State // Retrieving visible micro-models
\State Find visible micro-models for $POV_n$ in $vec\_vm$ 
\State // Caching
\For {each micro-model $m$ in $vec\_vm$}
    \If{$m$ in cache} \Comment{Hit}
        \State Continue
    \Else \Comment{Miss}
        \If{Cache is not full}
            \State Load $m$ to cache
        \Else
            \State Replace the LRU micro-model in cache with $m$
        \EndIf
    \EndIf
\EndFor
\State // Memory copy
\State Copy $vec\_vm$ from system RAM to GPU VRAM
\end{algorithmic}
\end{algorithm}

\subsubsection{Prefetching}
\begin{algorithm}[t]
\caption{Prefetching}\label{alg:prefetching}
\textbf{Input:} Sequence of previous of $POVs$.\\
\textbf{Output:} Cached micro-models after prefetching for $POV_{n+1}$.
\begin{algorithmic}[1]
\State // Retrieving visible micro-models
\State Predict visible micro-models for $POV_{n+1}$ in $vec\_vb\_next$
\State // Prefetching
\For {each micro-model $m$ in $vec\_vb\_next$}
    \If{$m$ not in cache}
        \If{$rendering\_done == false$} \Comment{Prefetching}
            \If{Cache is not full}
                \State Load $m$ to cache
            \Else
                \State Replace LRU micro-model in cache with $m$
            \EndIf
        \Else
            \State break \Comment{Stop prefetching (Rendering finished)}
        \EndIf
    \EndIf
\EndFor
\end{algorithmic}
\end{algorithm}

\begin{algorithm}[t]
\caption{Rendering}\label{alg:rendering}
\textbf{Input:} Visible micro-models.\\
\textbf{Output:} Visualization image.
\begin{algorithmic}[1]
\State $rendering\_done = false$
\For{$i \gets Row_{Start}$ to $Row_{End}$}
  \For{$j \gets Column_{Start}$ to $Column_{End}$}
    \For{$s \gets Sample_{First}$ to $Sample_{Last}$}
      \State $x, y, z = getSampleLocation(s)$
      \State $v = valueMicroModel(x, y, z)$
      \State $applyTFs(TF_{Color}, TF_{Opacity}, v, color)$
      \State $g_x, g_y, g_z = gradientMicroModel(x, y, z)$
      \State $c_{diffuse}, c_{specular} = lightCoefficient(g_x, g_y, g_z)$
      \State $addShading(c_{diffuse}, c_{specular}, color)$
      \State $O_{cumulative} = doColorCompositing(color)$
      \If{$O_{cumulative} > O_{Max}$} \Comment{Termination}
        \State break
      \EndIf
    \EndFor
    \State $setPixelColor(color)$
  \EndFor
\EndFor
\State $rendering\_done = true$
\end{algorithmic}
\end{algorithm}

Prefetching will predict the bag of visible micro-models for the next POV and update the same cache in RAM before the caching of the next POV so that the next caching will have a lower miss rate and therefore have less caching time. Prefetching is executed in parallel with rendering. To evaluate the performance of our framework, we considered three prefetching algorithms, APPA~\cite{7965175}, ForeCache~\cite{10.1145/2882903.2882919} and LSTM~\cite{8365978}, which are applicable to multi-resolution visualization systems for handling large-scale datasets. Prefetching involves two sub-steps as shown in Algorithm \ref{alg:prefetching}: 1) Predictive prefetching algorithm will predict the possible visible micro-models based on historical information from data (APPA) or user behavior (ForeCache, LSTM), where APPA predicts the parent POV (pPOV) while ForeCache and LSTM predict the next POV. 2) The prefetching step will update the system cache by loading the data of missing micro-models from storage by following the LRU replacement policy. The prefetching step needs to be preempted as soon as the rendering procedure is finished. This prefetching procedure is handled by another CPU thread in parallel with the CPU thread that handles caching and rendering. 

\subsubsection{Rendering}
We pick the typical direct volume rendering (DVR) using ray casting as an example of visualization task, however, other 3D volume visualizations using GPU are also applicable to leverage our proposed encoding and decoding solution to handle large-scale volumes. The rendering procedure is accelerated using a CUDA parallel programming model on GPU at the pixel level to generate the final visualization image. The rendering is a front-to-back ray casting algorithm with early termination as shown in Algorithm \ref{alg:rendering}. Functions $valueMicroModel(x, y, z)$ and $gradientMicroModel(x, y, z)$ are two key functions implemented in CUDA for querying value and gradient directly from Adaptive-FAM micro-models. Those functions are executed in parallel to accelerate the rendering time. On the other hand, the execution times of those querying functions grow sublinearly with respect to the NCP used to encode the micro-model. So we further decrease the rendering time by using a minimal suitable NCP while still maintaining the required error bound. The CUDA kernel function is called by the same CPU thread that handles caching. Synchronization between the two CPU threads is implemented using a flag named $rendering\_done$ to ensure that the prefetching only executes during the time of rendering.

\section{Experiments and evaluation}

\begin{figure}[t]
    \centering
    \begin{subfigure}[b]{0.31\linewidth}
        \centering
        \includegraphics[trim=0 0 0 0,clip,width=\linewidth]{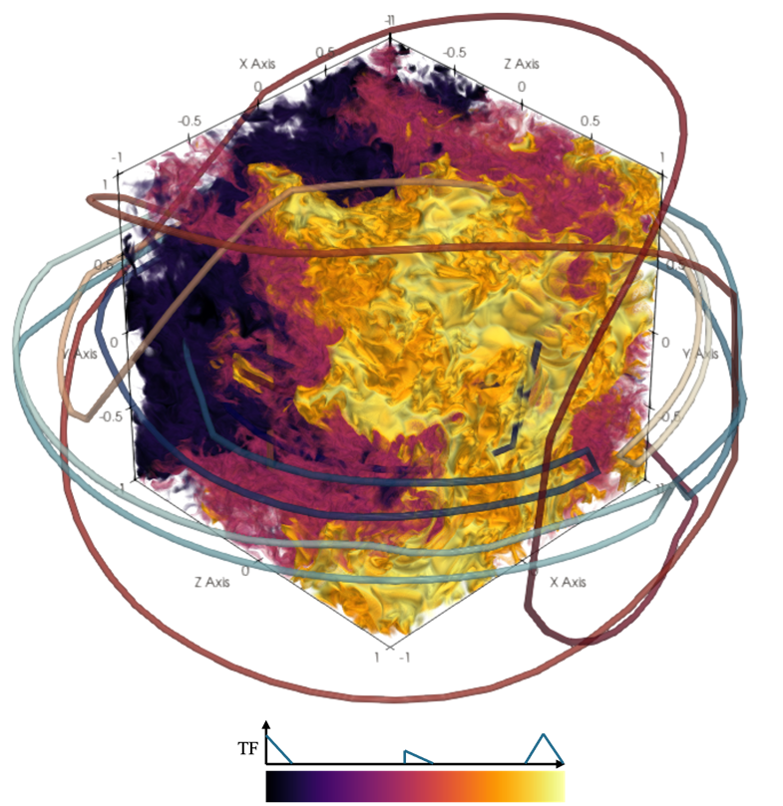}
        \caption{Rayleigh-Taylor}
        \label{fig:test_1_sequences}
    \end{subfigure}
    \begin{subfigure}[b]{0.31\linewidth}  
        \centering 
        \includegraphics[trim=0 0 0 0,clip,width=\linewidth]{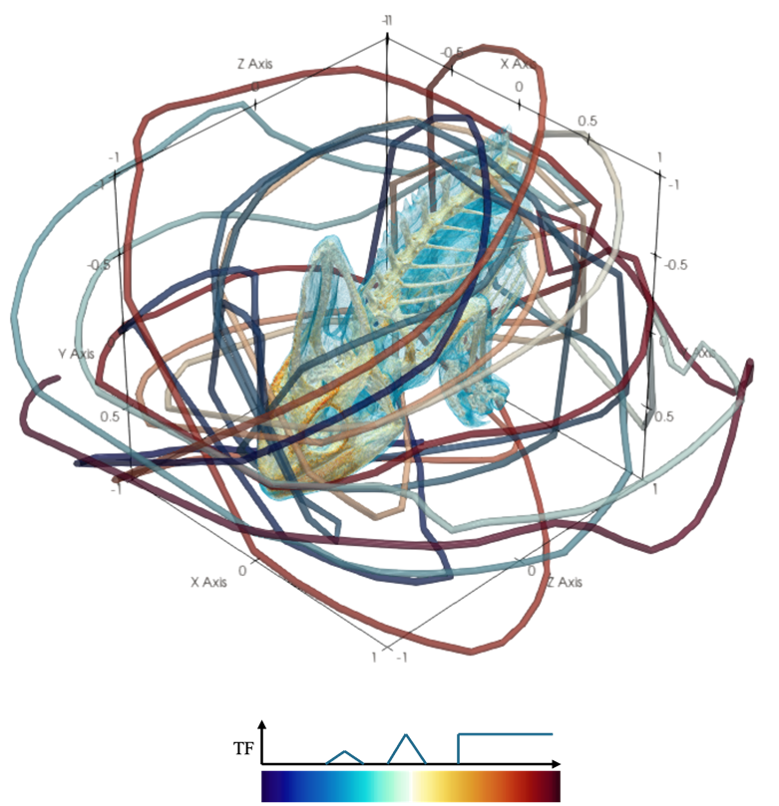}
        \caption{Chameleon}
        \label{fig:test_2_sequences}
    \end{subfigure}
    \begin{subfigure}[b]{0.31\linewidth}   
        \centering 
        \includegraphics[trim=0 0 0 0,clip,width=\linewidth]{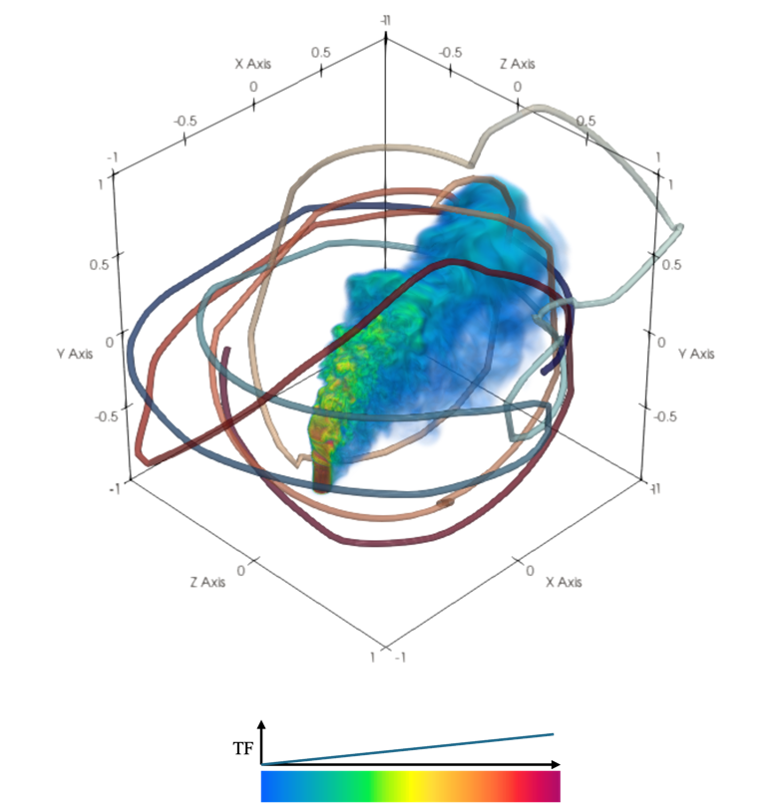}
        \caption{Flame}
        \label{fig:test_3_sequences}
    \end{subfigure}
    \vskip\baselineskip
    \begin{subfigure}[b]{0.31\linewidth}   
        \centering 
        \includegraphics[trim=0 0 0 0,clip,width=\linewidth]{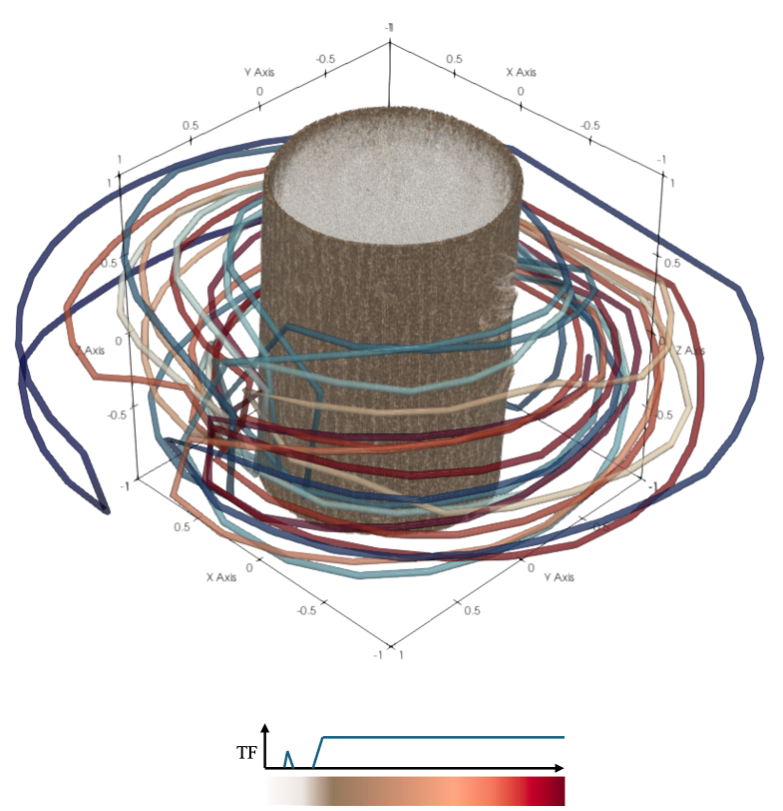}
        \caption{Branch}
        \label{fig:test_4_sequences}
    \end{subfigure}
    \begin{subfigure}[b]{0.31\linewidth}   
        \centering 
        \includegraphics[trim=0 0 0 0,clip,width=\linewidth]{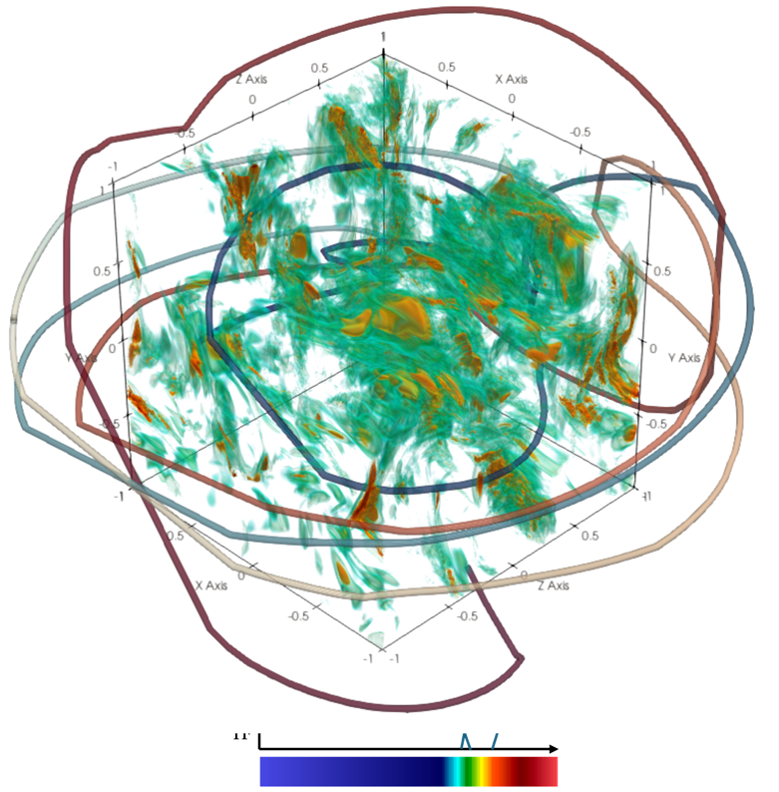}
        \caption{Rotstrat}
        \label{fig:test_4_sequences}
    \end{subfigure}
    \caption{Large-scale testing scientific datasets and their user exploratory trajectories.}
    \label{fig:test_datasets}
\end{figure}

\subsection{Dataset and Experimental Setup}
\subsubsection{Synthetic Dataset}
For rendering quality evaluation, we employ the continuous 3D function of Marschner-Lobb to provide a precise ground-truth value and gradient queries from arbitrary locations within the domain. The Marschner-Lobb function was originally introduced for assessing 3D resampling filters as they're applied to the traditional Cartesian cubic lattice\cite{346331}. Due to its complex amplitude distribution across various frequencies, accurately reconstructing a Marschner-Lobb signal from its discrete samples is challenging, making it an outstanding benchmark for evaluating reconstruction quality. We create a large-scale synthetic discrete dataset by sampling a dense regular grid from the 3D Marschner-Lobb function to serve as the input to Adaptive-FAM and other compressors applicable for multi-resolution. The Marschner-Lobb function used in this paper is defined as:

\begin{equation}
F(x, y, z)=\frac{1-\sin(\frac{\pi z}{2}) + \alpha(1+\rho_{r}(\sqrt{x^2+y^2}))}{2(1+\alpha)}
\end{equation}
where
\begin{equation*}
\rho_{r}(r) = \cos(2\pi f_{M} \cos(\frac{\pi r}{2}))
\end{equation*}
We use $f_{M} = 6$ and $\alpha = 0.05$ to generate a synthetic discrete Marschner-Lobb (ML) dataset with a resolution of $1200\times 1200\times 1200$. The spatial boundary on each dimension is $[0, 7]$. 

\subsubsection{Real Datasets}
For performance evaluation, we select 5 large-scale volume datasets with distinct spatial features collected from diverse domains. The Rayleigh-Taylor dataset is a time step of a density field in a simulation of the mixing transition in Rayleigh-Taylor instability. The Chameleon dataset is CT scan of a chameleon. The Flame dataset is a simulated combustion 3D scalar field. The Branch dataset is a microCT scan of dried wood branch. Rotstrat dataset is the temperature field of a direct numerical simulation of rotating stratified turbulence. \autoref{tab:datasets} shows the detailed information of the testing datasets. All datasets use float32 data type. Each dataset has a corresponding exploratory trajectory collected from a real human user containing 400 points of view. \autoref{fig:test_datasets} demonstrates the testing datasets and their user exploratory trajectories, together with the detailed color and opacity transfer functions (TFs) used for rendering.

\begin{table}[h]
  \caption{Real datasets information of resolution, size, micro-blocks resolution, and error bound (Root Mean Squared Error) for encoding. A lower error bound is applied to simpler datasets (Chameleon and Flame) to evaluate the maximum achievable compression ratios.}
  \label{tab:datasets}
  \scriptsize%
	\centering%
  \begin{adjustbox}{width=0.45\textwidth}
      \begin{tabu}{ c c c c c }
          \toprule
          Dataset & Resolution & Size & Micro-block Size & Error Bound (RMSE)\\
          \midrule
          Rayleigh-Taylor & $1024^3$           & 4 GB    & $64^3$         & 0.0001 \\
          \midrule
          Chameleon       & $1024^2\times1088$ & 4.25 GB & $64^2\times68$ & 0.00001 \\
          \midrule
          Flame           & $1200^3$           & 6.45 GB & $75^3$         & 0.00001 \\
          \midrule
          Branch          & $2048^3$           & 32 GB   & $128^3$        & 0.0001 \\
          \midrule
          Rotstrat        & $4096^3$           & 256 GB  & $256^3$        & 0.0001  \\
          \bottomrule
      \end{tabu}
  \end{adjustbox}
\end{table}

\begin{figure}[t]
 \centering 
 \includegraphics[trim=0 0 0 90,clip,width=0.985\columnwidth]{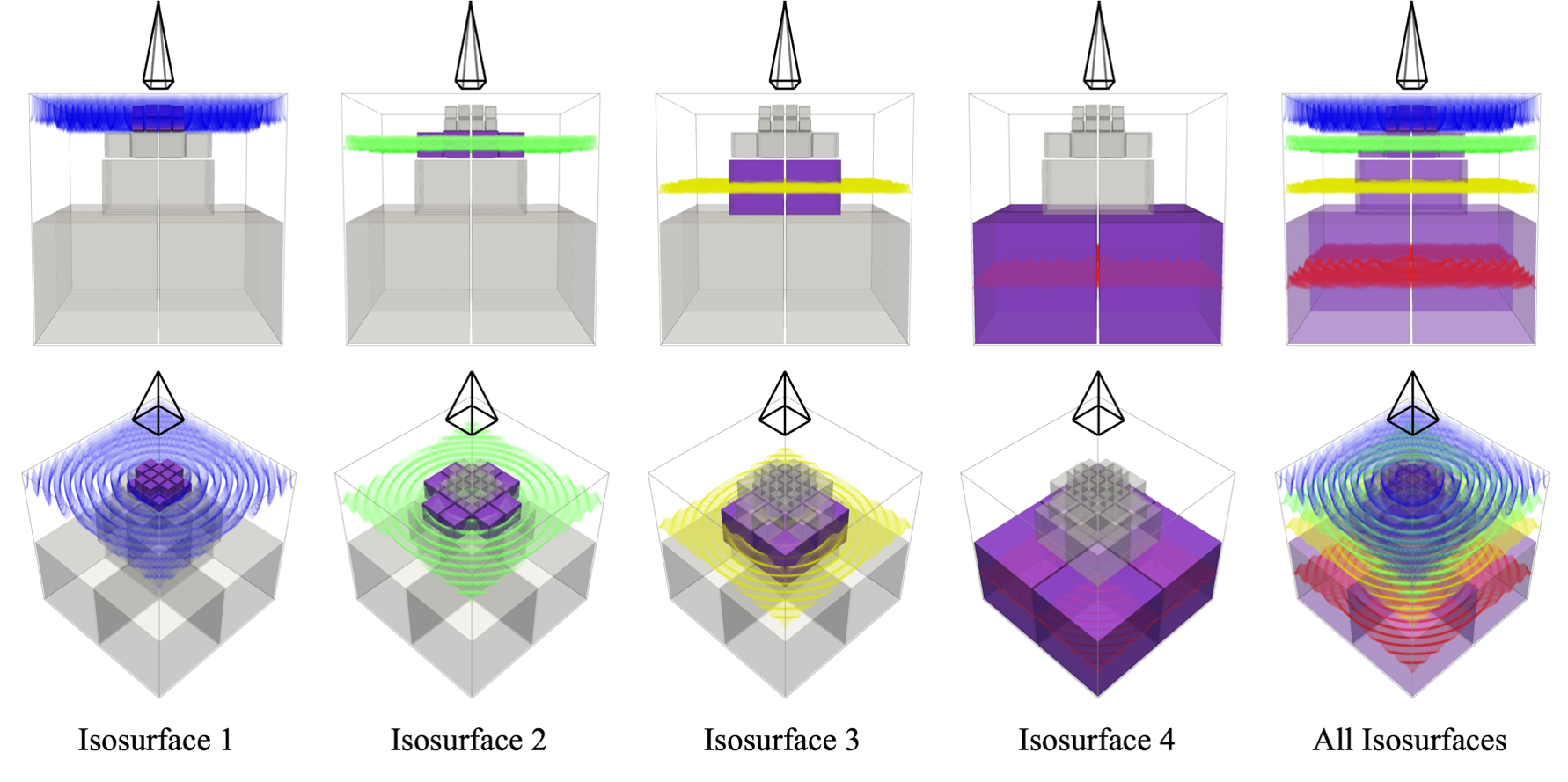}
 \caption{Testing POV and configurations of transfer functions for cross-level quality evaluation.}
 \label{fig:cross_eval}
\end{figure}

\subsubsection{Experimental Setup}
In the experiments, four LODs are employed by utilizing a bipartite strategy to partition micro-blocks for multi-resolution visualization. Four evenly distributed distance ranges ($[0, 0.8]$, $[0.8, 1.6]$, $[1.6, 2.4]$, $[2.4, \infty]$) are defined to determine the selection of micro-blocks for each LOD based on the distance between the user's point of view (POV) to the centroid of the micro-blocks. To ensure uniformity across spatial domains and multi-resolution policies, all volumes are rescaled to fall within the range of $[-1, 1]$ for the x, y, and z dimensions. The rendering performance scales linearly with the total number of pixels to render for all the rendering methods considered utilize pixel-level parallelization through CUDA. We select $512\times512$ as the image resolution for our experiments. A two-level memory hierarchy model, from storage to RAM, is used in the experiment for generalization. However, the benefit of our method can be easily scaled to multi-level memory hierarchies. The cache memory size is fixed as 200 micro-blocks for the multi-resolution rendering pipeline. To better demonstrate the effectiveness of all methods considered, we use a slow 4TB hard disk drive (HDD) with 5400 RPM on SATA interface as the lower level storage to simulate a worst-case scenario where high data movement latency between storage and RAM becomes the main bottleneck of the responsiveness. For testing the capability of our out-of-core method, we select an NVIDIA GTX 1080 Ti GPU with a VRAM of 8GB. 
The computing platform is a desktop featuring an Intel(R) Core(TM) i7-7700K CPU with 8 threads running at 4.20GHz, paired with 16GB of DDR4 DRAM clocked at 3200MHz, and operating on Ubuntu 20.04.4 LTS. Encoding is performed in parallel using CPU threads while rendering is accelerated by GPU. We disable the system cache of the operating system so that our measurements of caching time is more faithful to the real performance of the interactive visualization task.

\subsection{Rendering Quality Evaluation}
\begin{figure}[t]
 \centering 
 \includegraphics[width=0.8\columnwidth]{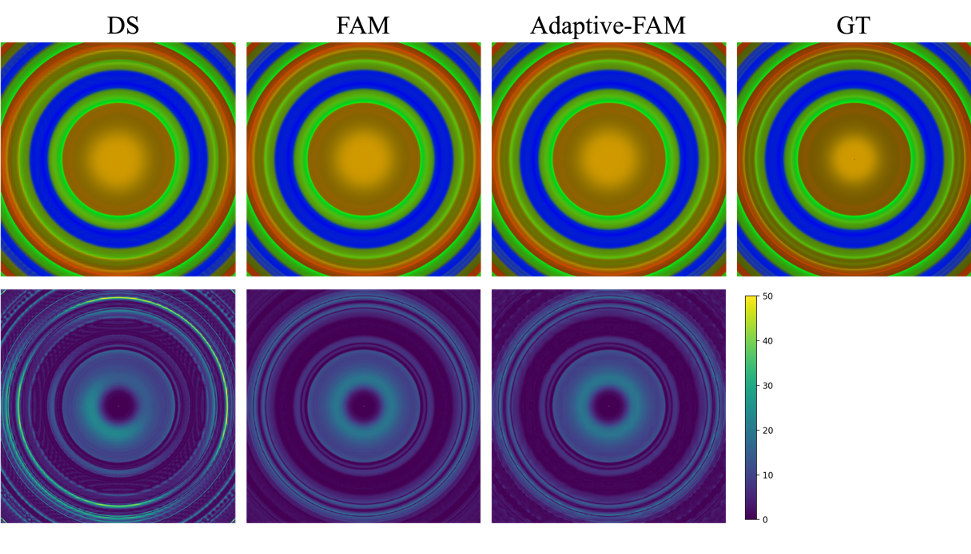}
 \caption{Rendering results of the cross-level quality evaluation on ML dataset using sample distance as 0.000025.}
 \label{fig:cross_images}
\end{figure}

\begin{figure}[t]
  \begin{subfigure}[b]{0.33\columnwidth}
    \includegraphics[trim=0 0 0 0,clip,width=\linewidth]{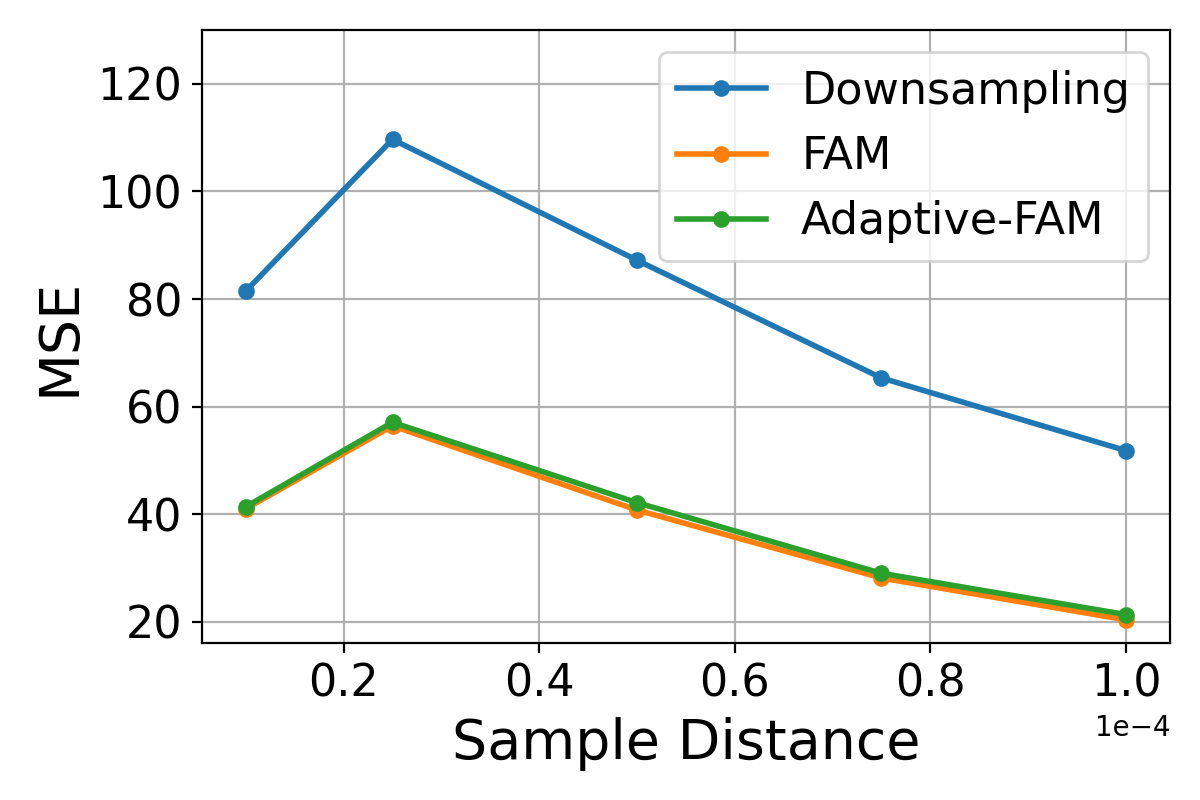}
    \caption{MSE}
  \end{subfigure}%
  \hspace*{\fill}   
  \begin{subfigure}[b]{0.33\columnwidth}
    \includegraphics[trim=0 0 0 0,clip,width=\linewidth]{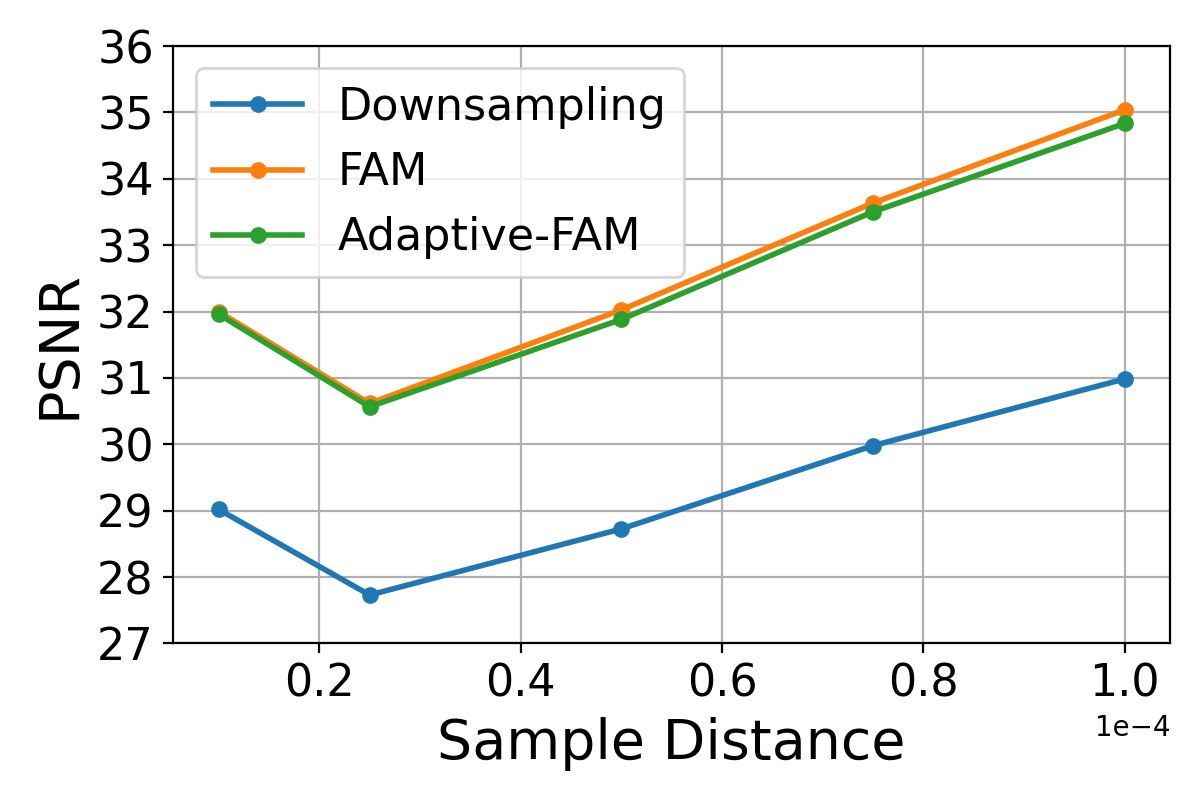}
    \caption{PSNR}
  \end{subfigure}%
  \hspace*{\fill}   
  \begin{subfigure}[b]{0.33\columnwidth}
    \includegraphics[trim=0 0 0 0,clip,width=\linewidth]{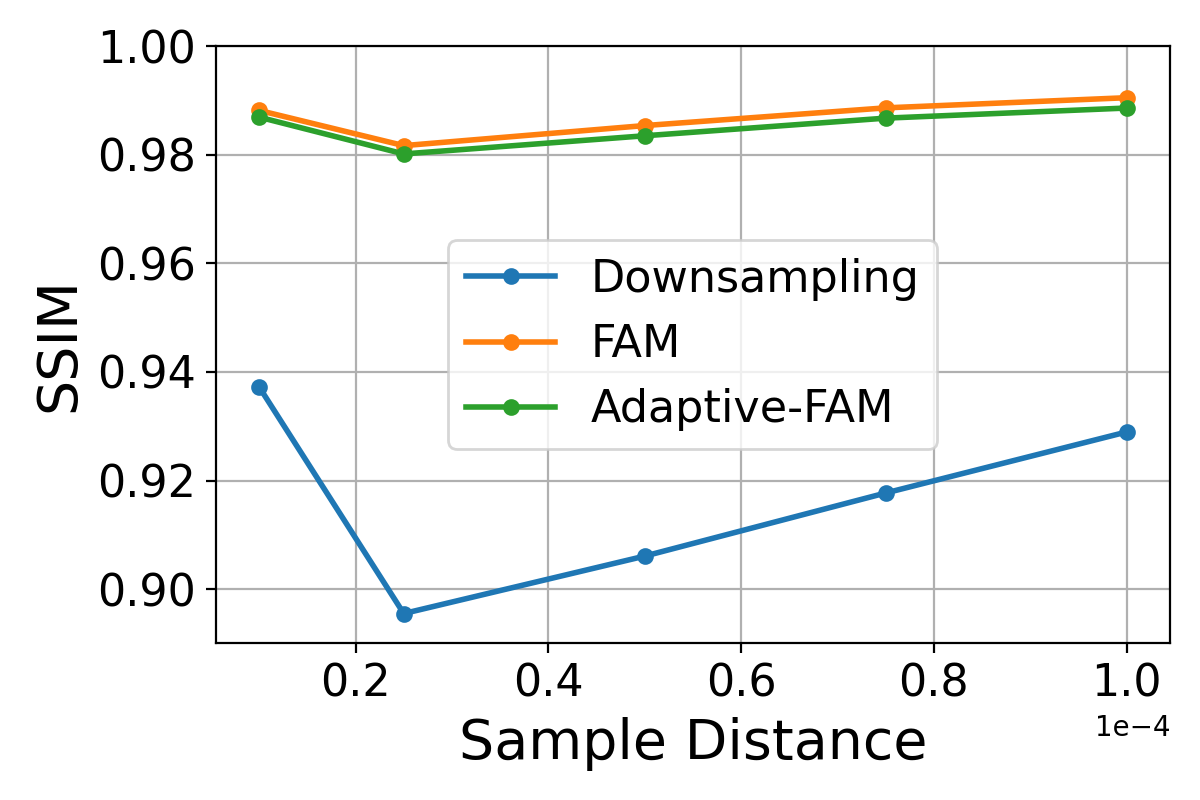}
    \caption{SSIM}
  \end{subfigure}%
\caption{Quantitative comparison of cross-level rendering quality of ML dataset.}
\label{fig:qualitative_image_metrics_cross}
\end{figure}

To evaluate the rendering quality of the proposed multi-resolution framework using proposed method, we practice a comparison of rendering results of large-scale synthetic ML data using various policies for constructing multi-resolution. we select a user POV whose visible region contains blocks from all 4 LODs at the same time. We also construct an isosurface crossing blocks of each LOD. The isosurfaces are set with transparency to make the user's POV see through each isosurface so that the final rendering results aggregate the rendering quality from all 4 LODs. We customize such visualization by adjusting the opacity and color transfer functions of the volume rendering. \autoref{fig:cross_eval} demonstrates the selected POV and data-dependent configuration of the testing ML dataset. We compare our Adaptive-FAM with a functional approximation encoding without adaptive encoding, FAM, as a baseline. This FAM model also utilizes our GPU-accelerated framework, so we can validate how Adaptive-FAM performs on both accuracy and rendering performance. We also consider the traditional down sampling methods (DS) as a reference. \autoref{fig:cross_images} shows the rendering images using down sampling, FAM, Adaptive-FAM, and the ground truth. We also tested several options for sample distance, one of the most fundamental factors of rendering quality, to reveal the trend of quality changes to such factors. As shown in \autoref{fig:qualitative_image_metrics_cross}, methods using functional approximation still give higher rendering accuracy than down sampling method. It is worth noticing that the Adaptive-FAM, although encoded uses smaller NCPs than FAM, has very close rendering accuracy to FAM. For encoding this large-scale ML dataset, the total micro-blocks modeled by FAM is around 7.7GB, while the total micro-blocks modeled by Adaptive-FAM is around 232MB. This means our Adaptive-FAM achieves similar rendering quality while greatly decreasing the sizes of the micro-models, which will help to decrease the caching time for more responsive input latency.

\begin{figure}[t]
 \centering 
 \includegraphics[width=1.0\columnwidth]{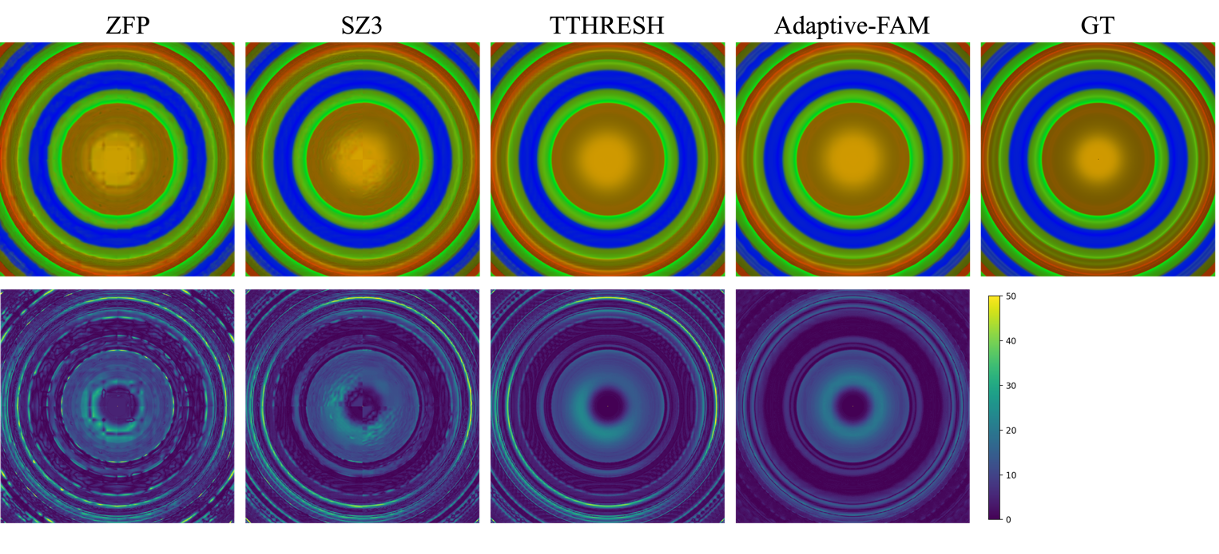}
 \caption{Rendering results using popular volume compressors and Adaptive-FAM on ML dataset.}
 \label{fig:compressors}
\end{figure}

\begin{figure}[t]
  \begin{subfigure}[b]{0.33\columnwidth}
    \includegraphics[trim=0 0 0 0,clip,width=\linewidth]{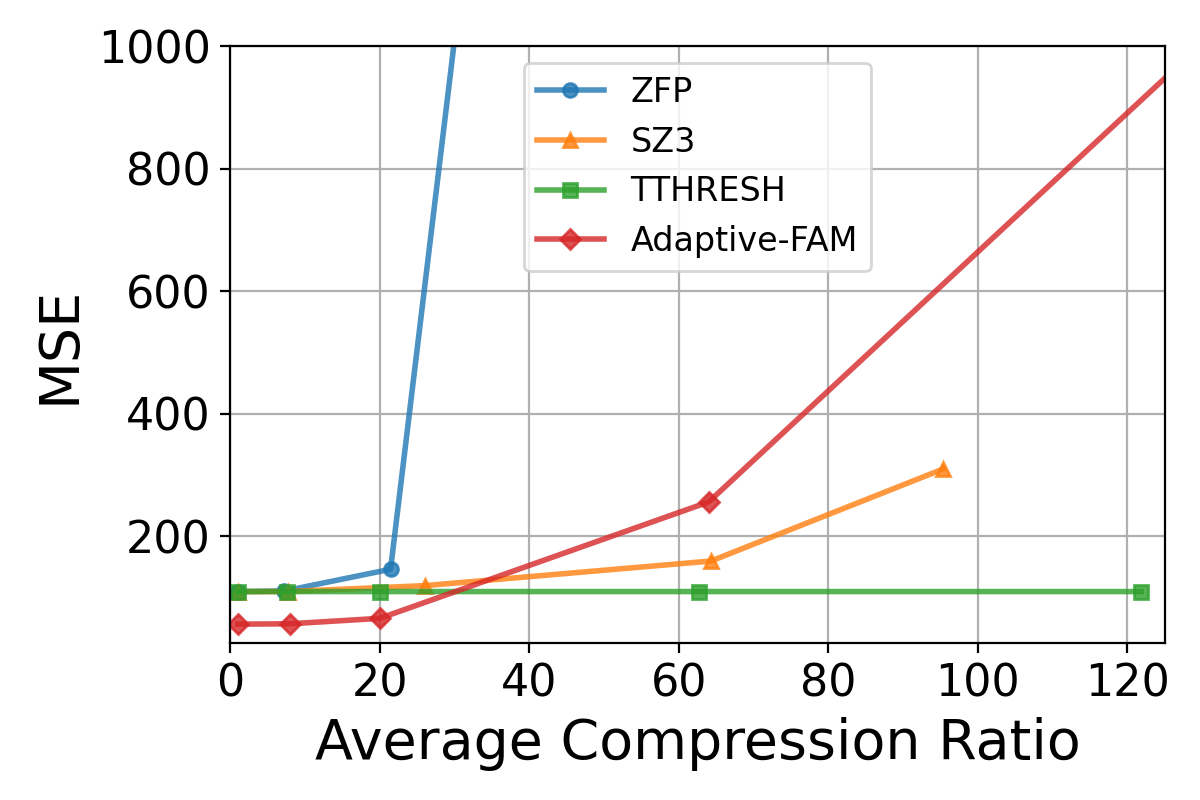}
    \caption{MSE}
  \end{subfigure}%
  \hspace*{\fill}   
  \begin{subfigure}[b]{0.33\columnwidth}
    \includegraphics[trim=0 0 0 0,clip,width=\linewidth]{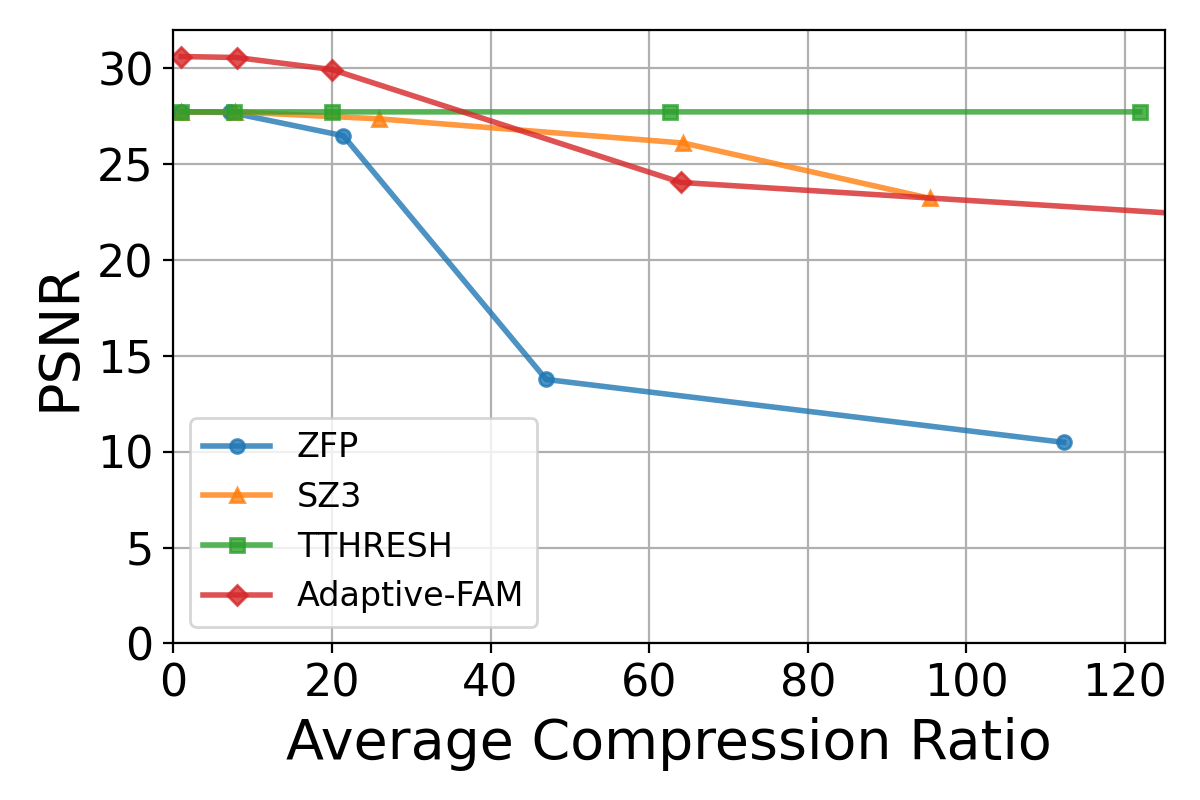}
    \caption{PSNR}
  \end{subfigure}%
  \hspace*{\fill}   
  \begin{subfigure}[b]{0.33\columnwidth}
    \includegraphics[trim=0 0 0 0,clip,width=\linewidth]{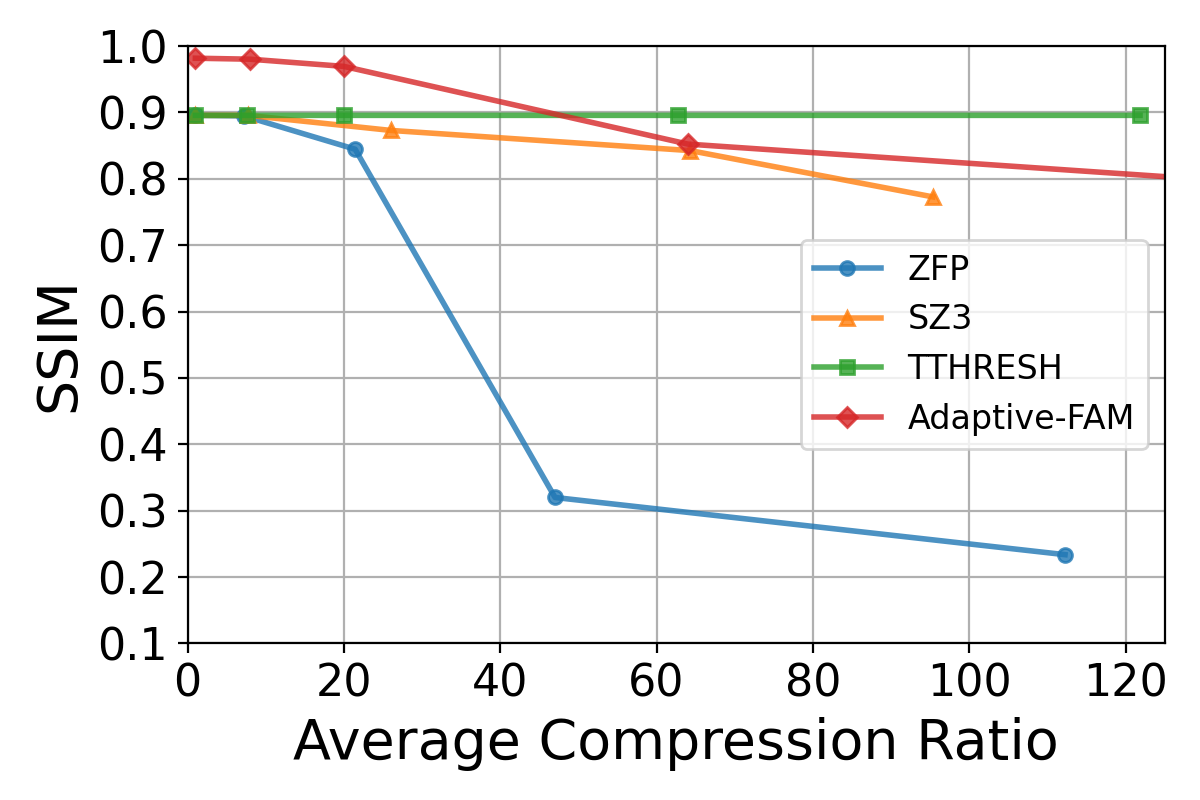}
    \caption{SSIM}
  \end{subfigure}%
\caption{Quantitative comparison using popular volume compressors and Adaptive-FAM.}
\label{fig:qualitative_compressors}
\end{figure}

The proposed Adaptive-FAM not only represents the data as a continuous high-order model but also does dynamic compression according to the content of the micro-blocks. We also introduce several popular volume compressors (ZFP, SZ3, and TTHRESH) to compare their rendering quality with Adaptive-FAM. All three volume compressors are configured to compress the ML dataset with a similar average compression ratio ($\approx20$) as Adaptive-FAM. All the compression methods encode the micro-blocks without including the ghost area. \autoref{fig:compressors} shows the rendering results of cross-level quality evaluation, where Adaptive-FAM produces the least amount of artifact. We can also clearly notice the artifact on the boundary of the neighboring blocks from ZFP and SZ3 due to its inaccurate reconstruction on the boundary sample. \autoref{fig:qualitative_compressors} shows the rendering accuracy as the average compression ratio changes. We can see that Adaptive-FAM gives the best rendering accuracy when the average compression ratio is less than 40, its accuracy falls behind the SZ3 and TThresh for aggressive compression. Although Adaptive-FAM is not an ideal volume compressor to handle a very high compression ratio, its random accessibility does not need a decompression process as other volume compressors do for querying. The decompression latency is also higher when the compression ratio is large, creating large overhead on caching for using those volume compressors as the micro-block encoders. Adaptive-FAM is more suitable for compressing micro-blocks in out-of-core multi-resolution applications.

\subsection{Performance Evaluation}
We evaluate the performance of our adaptive encoding and the proposed GPU-accelerated out-of-core multi-resolution framework using Adaptive-FAM. We evaluate the encoding time, caching time, rendering time, and overall input latency. The 5 real datasets together with their user exploratory trajectories are used to run the tests. Since there are 400 POVs for each trajectory, all measurements are averaged times used to render one frame.


\begin{table}[t]
  \caption{Encoding time using FAM and Adaptive-FAM on testing datasets. Adaptive-FAM representation is also compact with a practical compression ratio.}
  \label{tab:encoding_time}
  \scriptsize%
	\centering%
  \begin{adjustbox}{width=0.4\textwidth}
      \begin{tabu}{ c | c c | c  }
      \toprule
      Testing & \multicolumn{2}{|c|}{Encoding Time (Hours)} & Compression Ratio  \\
      Dataset &
      \multicolumn{2}{|c|}{FAM $\downarrow$  \ \ \ \ \ \ \ \  Adaptive-FAM $\downarrow$} &
      Adaptive-FAM $\uparrow$\\
      \midrule
      Rayleigh-Taylor & 1.14  & 0.76 & 15.71  \\
      \midrule
      Chameleon & 1.21 & 0.17 & 27.19   \\
      \midrule
      Flame & 2.16 & 0.26 & 35.31  \\
      \midrule
      Branch & 18.29 & 2.26 & 30.59  \\
      \midrule
      Rotstrat & 292.68 & 172.16 & 16.92  \\
      \bottomrule
      \end{tabu}
 \end{adjustbox}
\end{table}

\begin{figure}[t]
    \centering 
    \includegraphics[width=0.81\columnwidth]{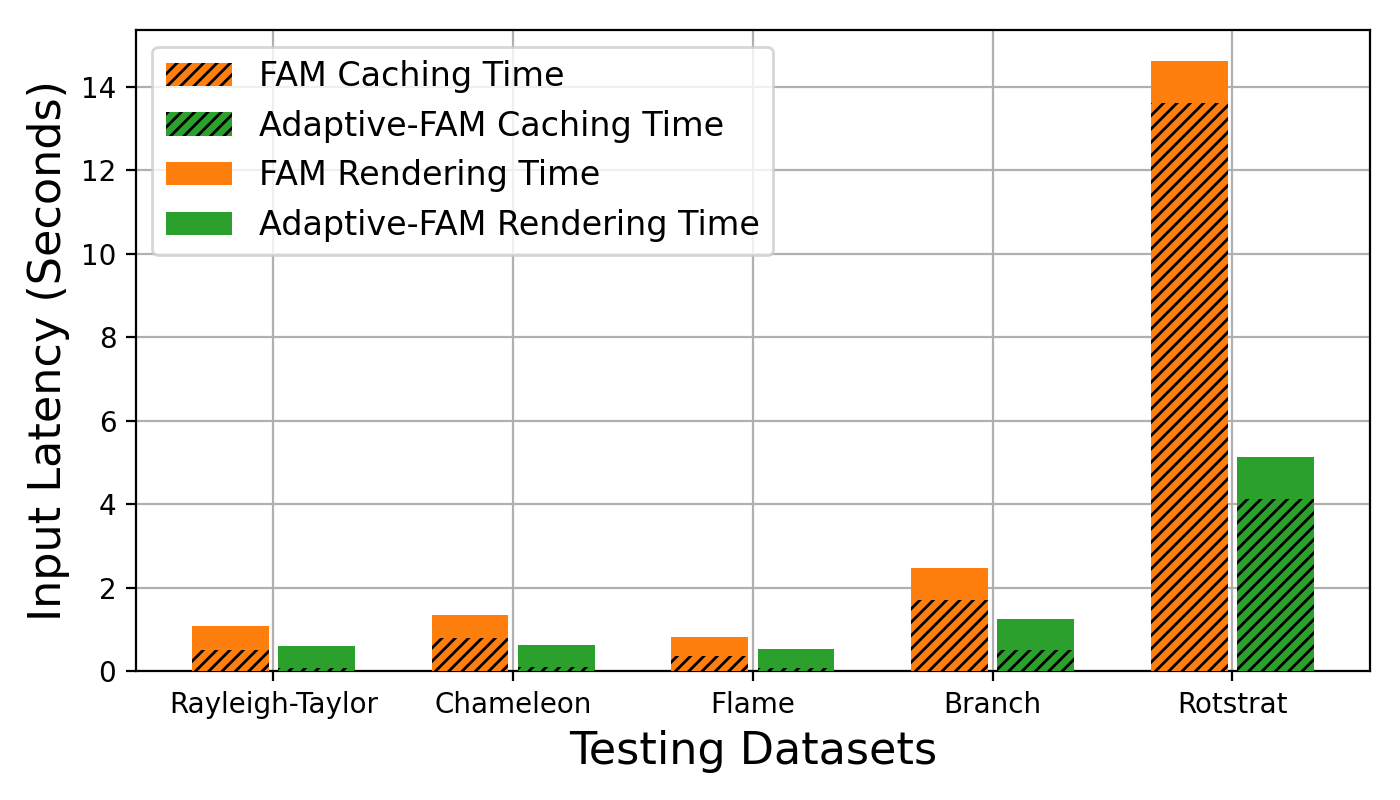}
    \caption{Input latency, consisted of caching and rendering time together, using FAM and Adaptive-FAM on testing datasets.}
    \label{fig:no_prefetching_performance}
\end{figure}

\subsubsection{Encoding Time}
\autoref{tab:encoding_time} shows the encoding time measurement for all 5 testing datasets. Our adaptive encoding greatly improves the encoding time compared with traditional functional approximation encoding for large-scale datasets. We can also observe that the encoding time saving depends on the complexity of the dataset. Our adaptive encoding works best for datasets with large and continuous empty spaces like the Chameleon, Flame, and Branch datasets. Adaptive-FAM representation is also compact which is beneficial to I/O intensive operations like caching and prefetching, which are crucial for efficiently managing large-scale datasets.

\subsubsection{Input Latency}
\begin{figure}[t]
    \centering
    \begin{subfigure}[b]{0.485\linewidth}
        \centering
        \includegraphics[trim=10 0 11 0,clip,width=\linewidth]{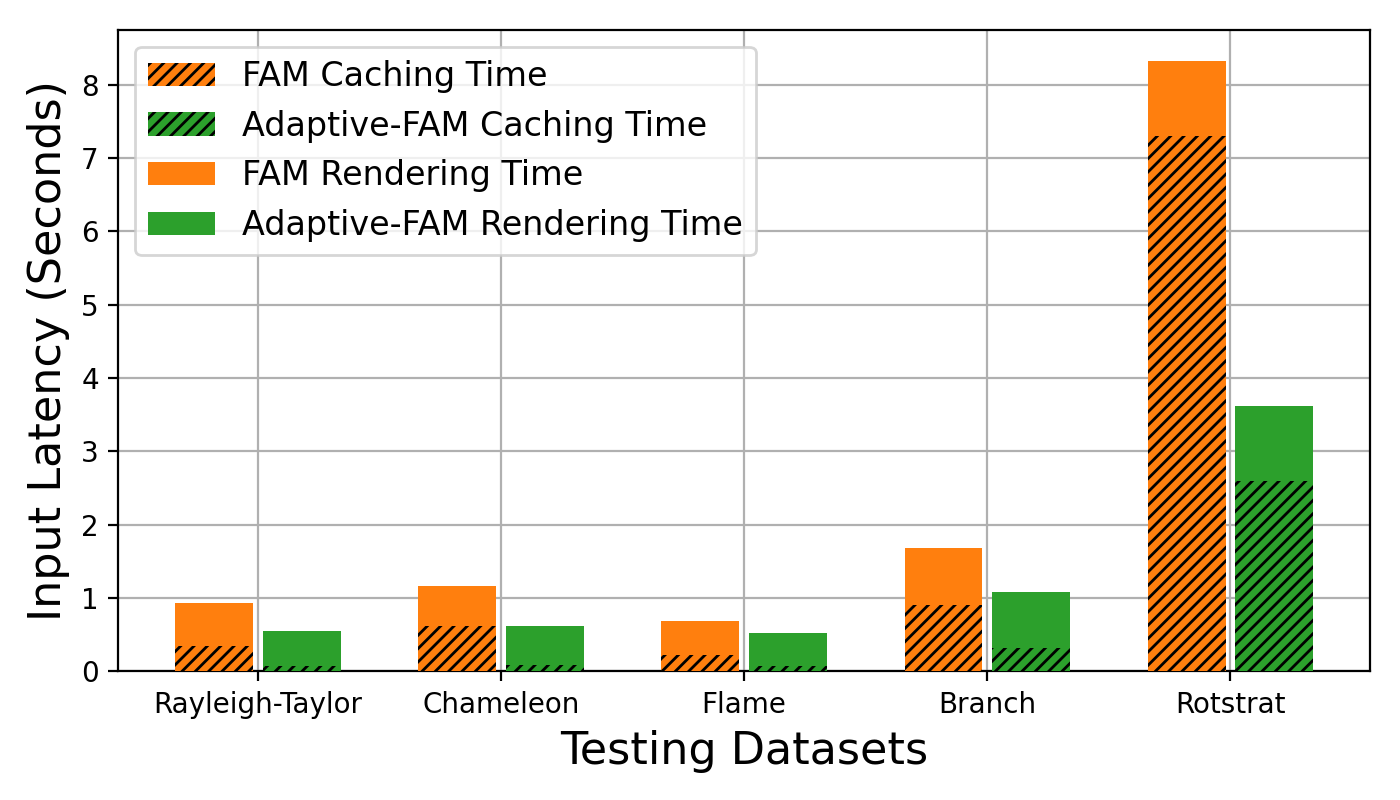}
        \caption{Latency with APPA}
        \label{fig:test_1_sequences}
    \end{subfigure}
    \hfill
    \begin{subfigure}[b]{0.485\linewidth}  
        \centering 
        \includegraphics[trim=10 0 11 0,clip,width=\linewidth]{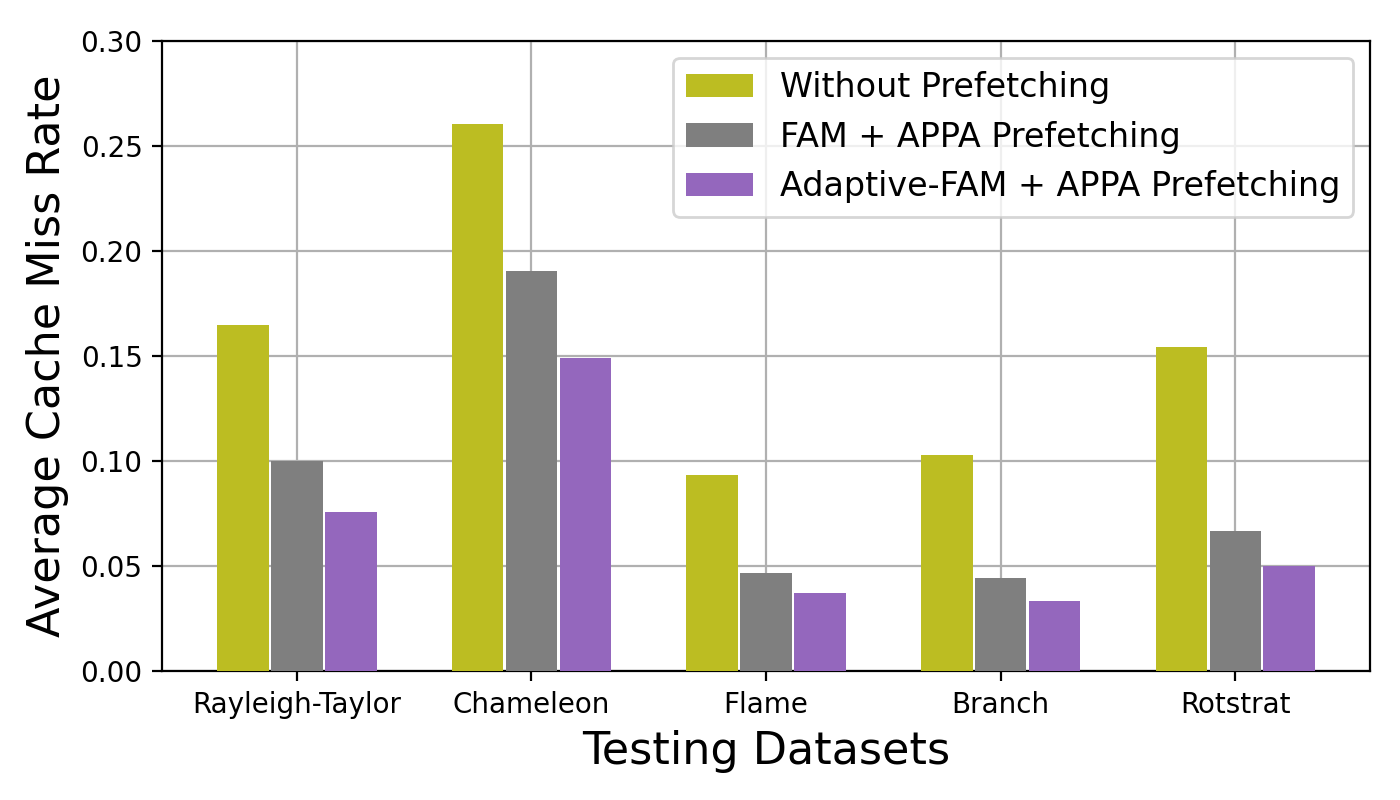}
        \caption{Caching with APPA}
        \label{fig:test_2_sequences}
    \end{subfigure}
    \vskip\baselineskip
    \begin{subfigure}[b]{0.485\linewidth}   
        \centering 
        \includegraphics[trim=10 0 11 0,clip,width=\linewidth]{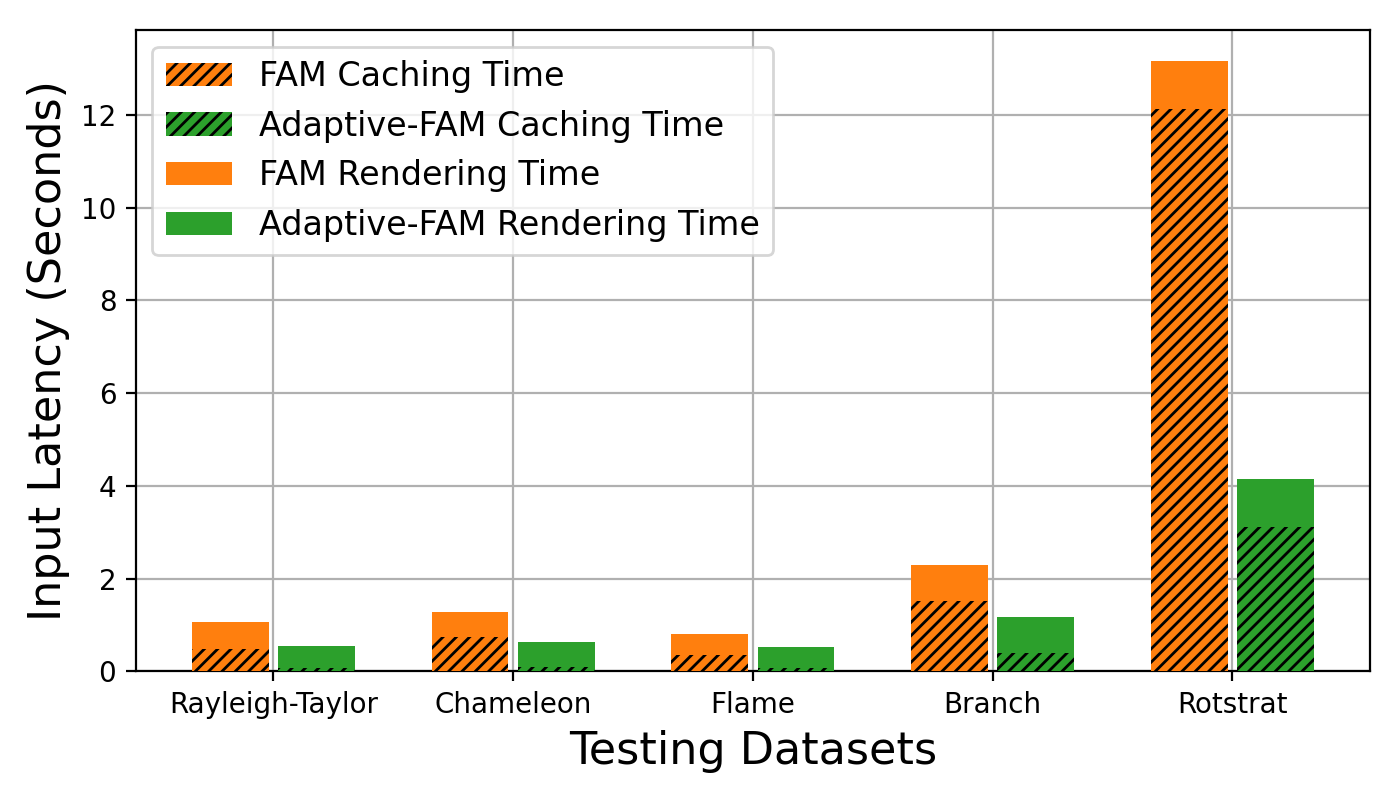}
        \caption{Latency with ForeCache}
        \label{fig:test_3_sequences}
    \end{subfigure}
    \hfill
    \begin{subfigure}[b]{0.485\linewidth}   
        \centering 
        \includegraphics[trim=10 0 11 0,clip,width=\linewidth]{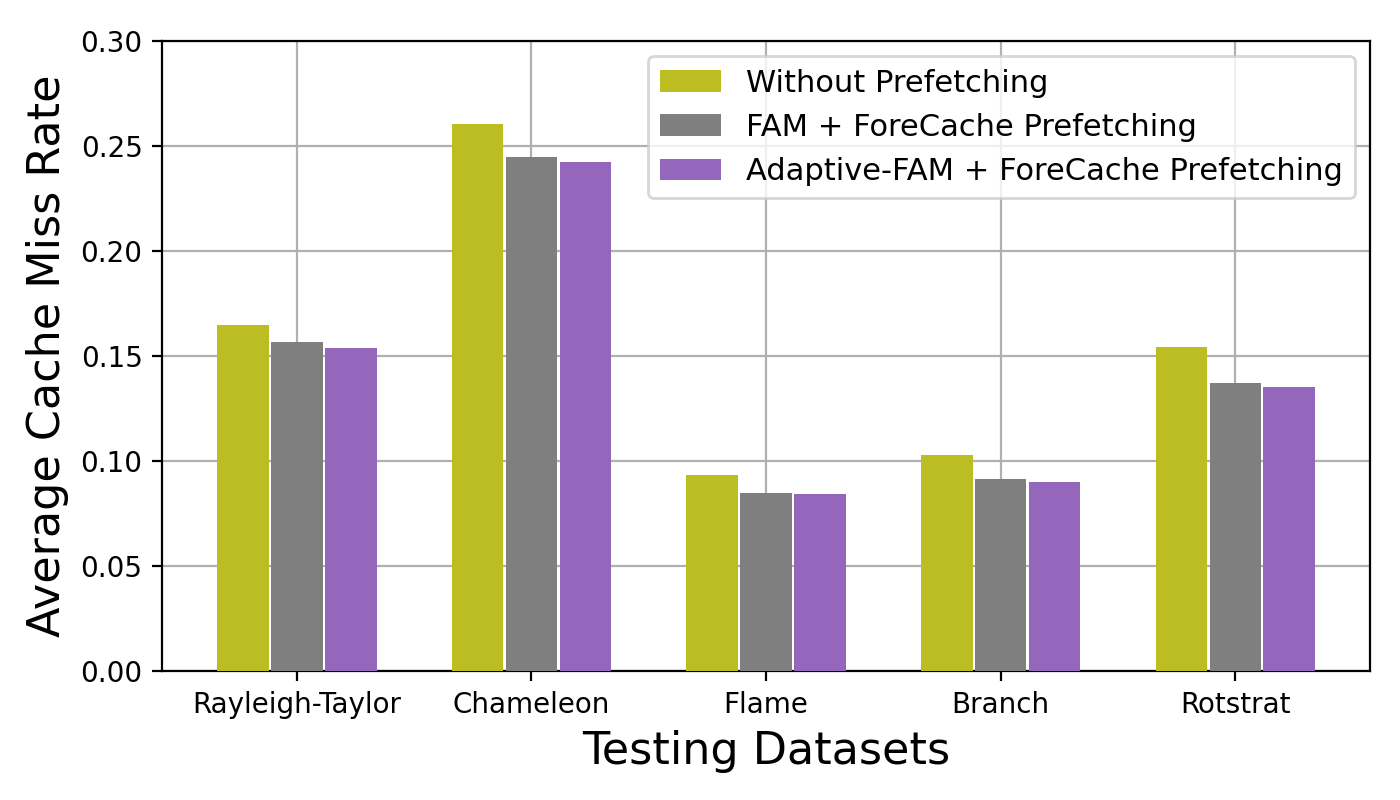}
        \caption{Caching with ForeCache}
        \label{fig:test_4_sequences}
    \end{subfigure}
    \vskip\baselineskip
    \begin{subfigure}[b]{0.485\linewidth}   
        \centering 
        \includegraphics[trim=10 0 11 0,clip,width=\linewidth]{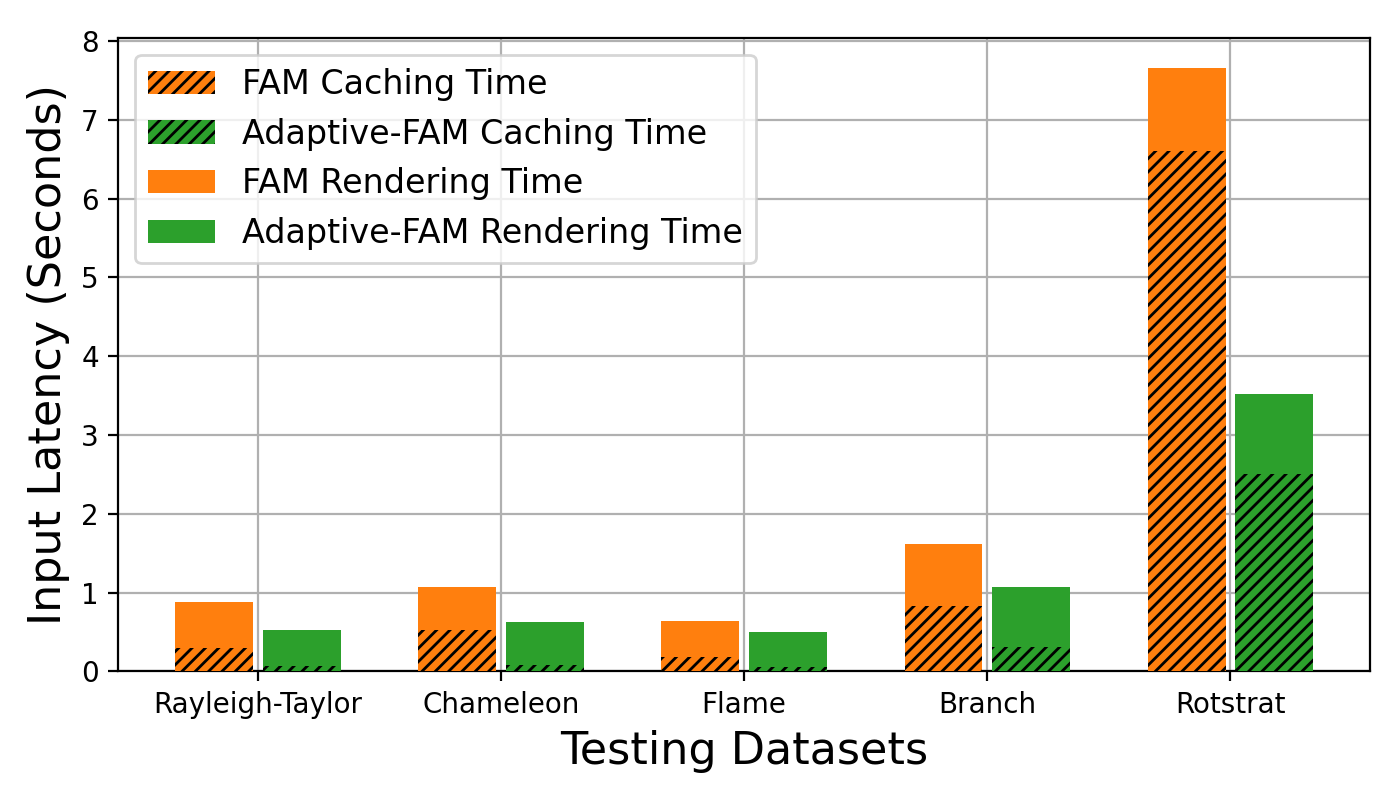}
        \caption{Latency with LSTM}
        \label{fig:test_3_sequences}
    \end{subfigure}
    \hfill
    \begin{subfigure}[b]{0.485\linewidth}   
        \centering 
        \includegraphics[trim=10 0 11 0,clip,width=\linewidth]{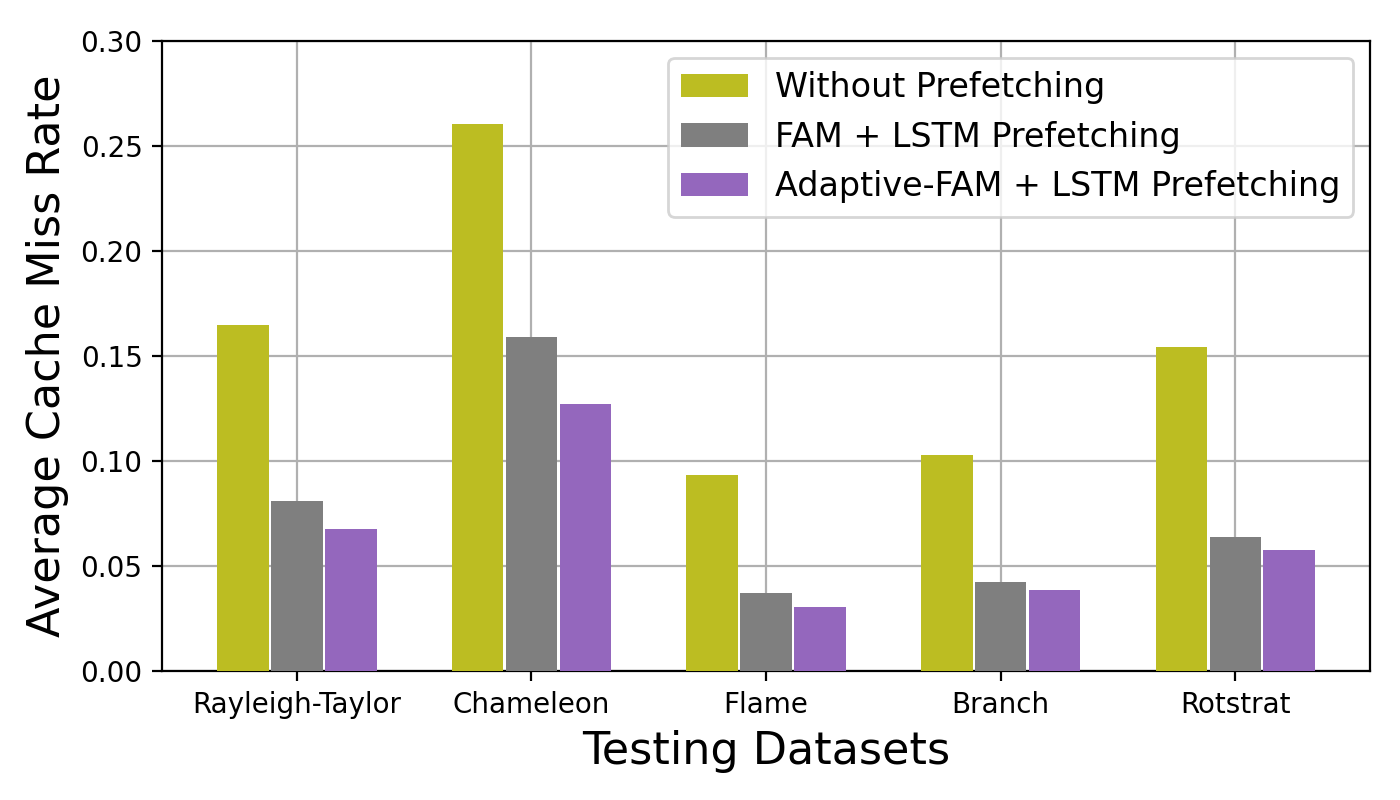}
        \caption{Caching with LSTM}
        \label{fig:test_4_sequences}
    \end{subfigure}
    \caption{Input latency and cache performance using various prefetching algorithms.}
    \label{fig:with_prefetching_performance}
\end{figure}

We first evaluate the performance without prefetching. \autoref{fig:no_prefetching_performance} shows the caching time, rendering time, and overall input latency on the 5 testing datasets using FAM and Adaptive-FAM. We can observe that Adaptive-FAM, because of the small size of micro-models using adaptive encoding, achieves a much shorter caching time than FAM. On the other hand, since the micro-models in Adaptive-FAM representation involve less average NCP, its rendering time is also faster than FAM. Since both FAM and Adaptive-FAM decode value and gradient from micro-model, whose querying time growing sublinearly with NCP, the rendering time is almost constant comparing to the growth of caching time as the data size increases. This validates why our Adaptive-FAM focuses on optimizing caching time by providing a more compact representation. For end-to-end time consumption, Adaptive-FAM can improve the overall input latency for all testing datasets. We then test our complete rendering framework with prefetching added, \autoref{fig:with_prefetching_performance} shows both the timing and the caching performance results. First, we can see that incorporating prefetching can improve the caching time for all testing datasets. This is achieved by decreased caching miss rate through prefetching as shown in the second column of \autoref{fig:with_prefetching_performance}. It is worth noticing that our Adaptive-FAM is more effective than FAM in decreasing the miss rate for all the prefetching algorithms. As a result, using prefetching can eventually improve the input latency. Second, various prefetching algorithms have different abilities to improve the caching miss rate. Third, prefetching does not change the result of which dataset Adaptive-FAM works best for. 

\begin{figure}[t]
    \centering 
    \includegraphics[width=0.81\columnwidth]{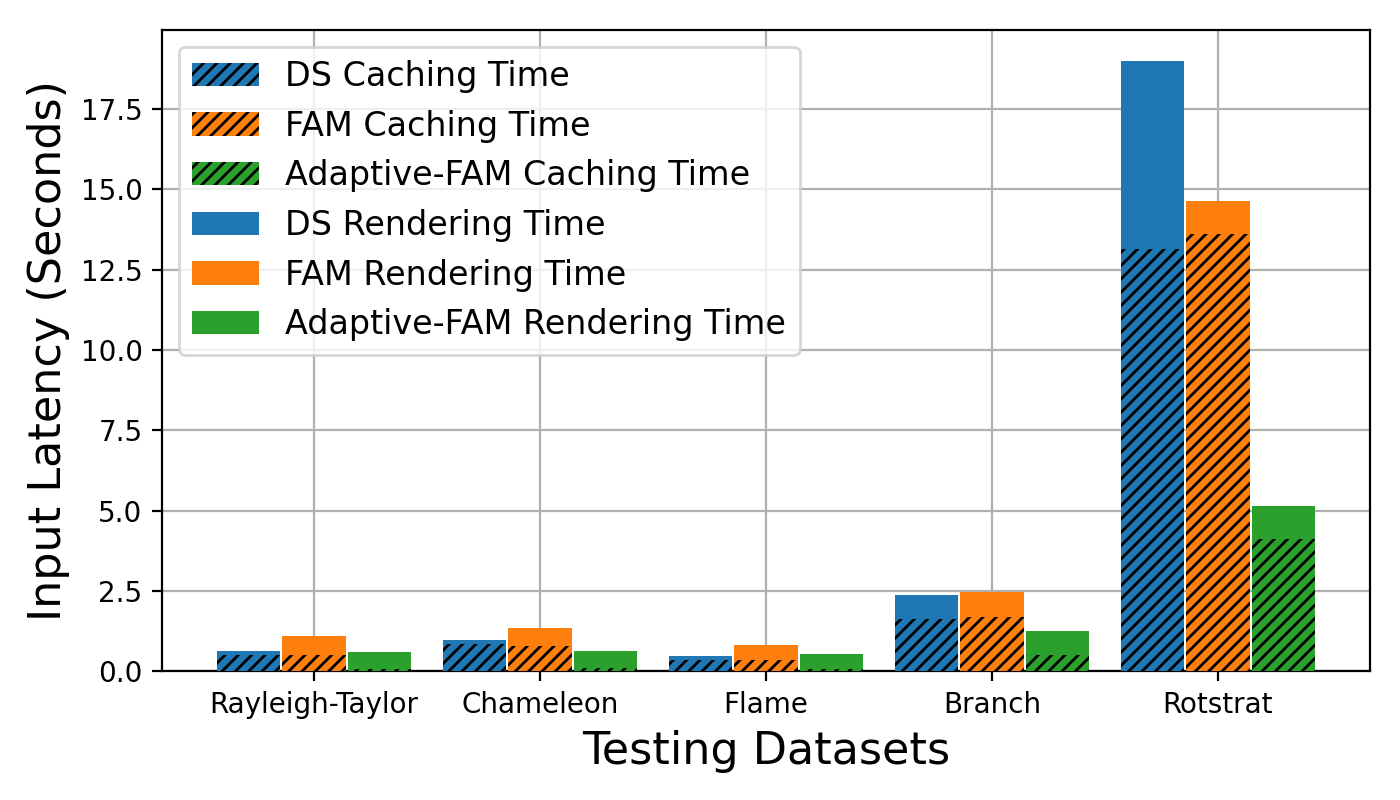}
    \caption{Caching time, rendering time, and overall input latency on 5 testing datasets using DS, FAM, and Adaptive-FAM.}
    \label{fig:no_prefetching_performance_all}
\end{figure}

We also compare the input latency of the proposed Adaptive-FAM with the state-of-the-art multi-resolution framework through down sampling (DS) utilizing GPU 3D texture for fast trilinear interpolation. \autoref{fig:no_prefetching_performance_all} shows the input latency using DS, FAM, and Adaptive-FAM. It can be seen that Adaptive-FAM is the fastest method for all the testing datasets except for the Flame dataset where DS is slightly faster. With the optimization on micro-model sizes, Adaptive-FAM is always faster than FAM by saving on I/O time during caching. DS is not performing well when handling large-scale datasets for two main reasons: First, caching time becomes the main bottleneck when the number of micro-blocks to load become larger. Second, although the query using GPU-accelerated 3D texture is already fast, it scales linearly with respect to the micro-block size. As a result, DS struggles when rendering extremely large datasets. On the other hand, Adaptive-FAM is designed to minimize caching time across memory hierarchy through a data-dependent compact representation. Moreover, the sublinear nature of decoding time with respect to micro-block size also helps to maintain a fast rendering procedure. In summary, Adaptive-FAM performs better when visualizing large-scale volumetric datasets with responsive input latency.


\section{Conclusion}
In this paper, we proposed a novel functional approximation multi-resolution representation, Adaptive-FAM, and an adaptive encoding method to efficiently encode a large number of micro-blocks partitioned from a large-scale dataset. A GPU-accelerated out-of-core multi-resolution framework is proposed to directly render visualization results from the representation with improved input latency compared with the traditional function approximation model. Our solution delivers superior rendering quality while maintaining responsiveness comparable to the already fast down sampling method leveraging GPU 3D texture, and performs better when handling large-scale datasets. The limitation of this work is that the benefit of encoding time is reduced when the input dataset values are highly dynamic throughout the entire volumetric space. In the future, we would like to explore other micro-block encoding methods leveraging implicit neural representations and incorporating multiple deep learning models into a unified multi-resolution framework.
\section*{Acknowledgement}
This research has been sponsored in part by the National Science Foundation grants IIS-1423487 and IIS-1652846 and Advanced Scientific Computing Research, Office of Science, U.S. Department of Energy, under contracts DE-AC02-06CH11357, program manager Margaret Lentz.




\bibliographystyle{abbrv-doi}

\bibliography{template}
\end{document}